\documentclass[useAMS,usenatbib,usegraphicx]{mn2e}
\usepackage{amssymb}
\usepackage{graphicx}

\graphicspath{{./}}
\newcommand{\spc}{\mathbf{x}}

\newcommand{\msun}{${\rm M}_{\sun}$}
\newcommand{\ms}{M_{\sun}}
\newcommand{\Ms}{M_{\sun}}
\newcommand{\nub}{\bar{\nu}}
\newcommand{\nue}{\nu_e}
\newcommand{\nueb}{\bar{\nu}_e}

\newcommand{\numt}{\nu_{\mu,\tau}}

\newcommand{\be}{\begin{equation}}
\newcommand{\ee}{\end{equation}}

\newcommand{\refsec}[1]{Sec.~\ref{#1}}
\newcommand{\refeq}[1]{Eq.~(\ref{#1})}

\newcommand{\reffig}[1]{Fig.~\ref{#1}}
\newcommand{\reftab}[1]{Table~\ref{#1}}

\newcommand{\apj}{ApJ}
\newcommand{\apjs}{ApJS}
\newcommand{\apjl}{ApJL}
\newcommand{\mnras}{MNRAS}
\newcommand{\prd}{PhRD}
\newcommand{\nat}{Nature}
\newcommand{\aap}{A\&A}
\newcommand{\araa}{ARA\&A}
\newcommand{\physrep}{Phys. Rep}

\usepackage{color}

\title[$\nu$-driven winds from NS merger remnants]{Neutrino-driven winds from neutron star merger remnants}

\author[A. Perego et al.]{A. Perego$^{1,2}$\thanks{E-mail:
albino.perego@physik.tu-darmstadt.de}, S. Rosswog$^3$, R. M. Cabez\'{o}n$^2$, O. Korobkin$^3$, R. K\"{a}ppeli$^4$, 
\newauthor A. Arcones$^1$, M. Liebend\"{o}rfer$^2$\\
$^{1}$Institut f\"ur Kernphysik, Technische Universit\"at Darmstadt, Schlossgartenstra{\ss}e 2, D-64289 Darmstadt, Germany\\
$^{2}$Physics Department, University of Basel, Klingelbergstrasse 82, CH-4056 Basel, Switzerland\\
$^{3}$The Oskar Klein Centre, Department of Astronomy, AlbaNova, Stockholm University, SE-106 91 Stockholm, Sweden\\
$^{4}$Seminar for applied Mathematics, ETH Z\"urich, R\"amistrasse 101, 8092 Z\"urich, Switzerland}

\begin{document}

\date{Submitted to MNRAS}

\pagerange{\pageref{firstpage}--\pageref{lastpage}} \pubyear{year}

\maketitle

\label{firstpage}

\begin{abstract}
We present a detailed, three-dimensional hydrodynamic study of the neutrino-driven winds 
that emerge from the remnant of a neutron star merger. Our simulations are performed 
with the Newtonian, Eulerian code FISH, augmented by a detailed, spectral neutrino leakage
scheme that accounts for heating due to neutrino absorption in optically thin conditions. 
Consistent with the earlier, two-dimensional study of Dessart et al. (2009), we find that 
a strong baryonic wind is blown out along the original binary rotation axis within $\approx$100 
milliseconds after the merger.
We compute a lower limit on the expelled mass of 
$3.5 \times 10^{-3}$ \msun, large enough to be relevant for heavy element nucleosynthesis.
The physical properties vary significantly between different wind regions.
For example, due to stronger neutrino irradiation, the polar regions show substantially 
larger electron fractions than those at lower latitudes.
This has its bearings on the nucleosynthesis: the polar ejecta produce interesting r-process 
contributions from $A\approx 80$ to about 130, while the more neutron-rich, lower-latitude 
parts produce in addition also elements up to the third r-process peak near $A\approx 195$. 
We also calculate the properties of electromagnetic transients that are powered by the 
radioactivity in the wind, in addition to the  ``macronova'' transient that
stems from the dynamic ejecta.
The high-latitude (polar) regions produce UV/optical transients reaching luminosities up to 
$10^{41} {\rm erg \, s^{-1}}$, which peak around 1 day in optical and 0.3 days in bolometric luminosity. The lower-latitude regions, due to their contamination with high-opacity heavy elements, produce dimmer and more red signals, peaking after $\sim 2$ days in optical and infrared.
Our numerical experiments indicate that it will be difficult to infer the collapse time-scale of the 
hypermassive neutron star to a black hole based on the wind electromagnetic transient, at least for collapse time-scales
larger than the wind production time-scale.
\end{abstract}


%
%
\section{Introduction}

\begin{figure*}
\begin{center}
\includegraphics[width = \linewidth]{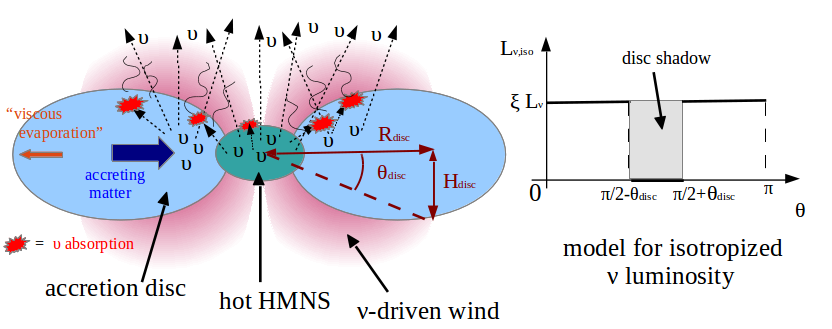}
\end{center}
\caption{Left: sketch of the neutrino-driven wind from the remnant of a BNS merger. 
The hot hypermassive neutron star (HMNS) and the accretion 
disc emit neutrinos, preferentially along the polar direction and at intermediate latitudes. A fraction
of the neutrinos is absorbed by the disc and can lift matter out of its gravitational potential.
On the viscous time-scale, matter is also ejected along the equatorial direction. Right: sketch of the 
isotropised $\nu$ luminosity we are using for our analytical estimates (see the main text for details).}
\label{fig: cartoon}
\end{figure*}

Neutron star mergers play a key role for several branches of modern astrophysics. They are --together
with neutron star-black hole coalescences-- the major astrophysical target of the ground-based gravitational wave detector facilities such as LIGO, VIRGO and KAGRA \citep{Acernese2008,Abbott2009,Harry2010,Somiya2012}. 
Moreover, such compact binary mergers have been among the very early suggestions
for the central engines of short gamma-ray bursts (sGRBs) \citep{Paczynski1986,Goodman1986,Eichler1989,Narayan1992}. 
While long GRBs (durations $>2$s) very likely have a different origin, compact binary mergers are the
most widely accepted engine for the category of short bursts (sGRBs). Over the years, however, contending models
have emerged and the confrontation of the properties expected from compact binary mergers with those observed
in sGRBs is not completely free of tension \citep[see][for recent reviews]{Piran2004,Lee2007,Nakar2007,Gehrels2009,Berger2011,Berger2013b}.
A binary neutron star merger (hereafter, BNS merger) forms initially a central,
hypermassive neutron star (HMNS) surrounded by a thick accretion disc.
During the merger
process a small fraction of the total mass becomes ejected via gravitational torques and hydrodynamic
processes (``dynamic ejecta''). The decompression of this initially cold and extremely neutron-rich
nuclear matter had long been suspected to provide favourable conditions for the formation 
of heavy elements through the rapid neutron capture process (the ``r-process'')  
\citep{Lattimer1974,Lattimer1976,Lattimer1977,Symbalisty1982,Eichler1989,Meyer1989,Davies1994}.  While initially only considered as an 
``exotic'' or second-best model behind core-collapse supernovae, there is nowadays a large literature 
that --based on hydrodynamical and nucleosynthetic calculations-- consistently finds that the dynamic 
ejecta of a neutron star merger is an extremely promising site for the formation of the heaviest elements 
with $A > 130$
\citep[see, e.g., ][]{Rosswog1999,Freiburghaus1999,Oechslin2007a,Metzger2010b,Roberts2011,
Goriely2011a,Goriely2011b,Korobkin2012,Bauswein2013a,
Hotokezaka2013,Kyutoku2013,Wanajo2014}. Core-collapse supernovae, on the contrary, seem seriously challenged 
in generating the conditions that are needed to produce elements with $A > 90$ 
\citep{Arcones2007,Roberts2010,Fischer2010,Huedepohl2010}.
A possible exception, though, may be magnetically driven explosions of rapidly rotating stars \citep{Winteler2012,Moesta2014}. 
Such explosions, however, require a combination of rather extreme properties of the pre-explosion star and
are therefore likely rare.\\
Most recently, the idea that compact binary mergers are related to both sGRBs and the nucleosynthesis of the heaviest
elements has gained substantial observational support. In June 2013, the SWIFT satellite detected a 
relatively nearby ($z=0.356$) sGRB, GRB130603B, \citep{Melandri2013} for which the Hubble Space Telescope
\citep{Tanvir2013,Berger2013a} detected a nIR point source, 9 days after the burst.
The properties of this second detection are close to model predictions 
\citep{Kasen2013,Barnes2013a,Tanaka2013a,Grossman2014,Rosswog2014a,Tanaka2014a} 
for the so-called ``macro-'' or ``kilonovae'' \citep{Li1998,Kulkarni2005,Rosswog2005,Metzger2010a,Metzger2010b,Roberts2011},
radioactively powered transients from the decay of freshly produced r-process elements. In particular, the delay of several days
between the sGRB and the nIR detection is consistent with the expanding material having very large opacities, as predicted for
very heavy r-process elements \citep{Kasen2013}.
If this interpretation is correct, GRB130603B would
 provide the first observational confirmation of the long-suspected link between compact binary mergers, 
heavy elements nucleosynthesis and gamma-ray bursts.\\
There are at least two more channels, apart from the dynamic ejecta, by which a compact binary merger releases matter into space,
and both of them are potentially interesting for nucleosynthesis and --if enough long-lived
radioactive material is produced-- they may also power additional electromagnetic transients. 
The first channel is the post-merger accretion disc. As it evolves viscously, expands and cools, the 
initially completely dissociated matter recombines into alpha-particles and --together with viscous heating-- 
releases enough energy  to unbind an amount of material that is comparable 
to the dynamic ejecta \citep{Metzger2008,Beloborodov2008,Metzger2009b,Lee2009,Fernandez2013}.\\
The second additional channel is related to  neutrino-driven winds, the basic mechanisms of which are
sketched in \reffig{fig: cartoon}.
This wind is, in several respects, similar to the one that emerges
from proto-neutron stars. In particular, in both cases a similar amount of gravitational binding energy 
is released over a comparable (neutrino diffusion) time-scale, which results in a luminosity of
$L_\nu \sim \Delta E_{\rm grav}/\tau_{\rm diff} \sim 10^{53}$ erg/s and neutrinos 
with energies $\sim 10-15$ MeV.
Under these conditions, energy deposition due to neutrino absorption is likely to unbind a fraction of the
merger remnant. In contrast to proto-neutron stars, however, the starting point is extremely neutron-rich
nuclear matter, rather than a deleptonizing stellar core. At remnant temperatures of several MeV, 
electron anti-neutrinos dominate over electron neutrinos, contrary to the proto-neutron
star case. Based on scaling relations from the proto-neutron star context \citep{Duncan1986,Qian1996}, early 
investigations discussed neutrino-driven winds from merger remnants either in an order-of-magnitude sense 
or via parametrized models \citep{Ruffert1997,Rosswog2002b,Rosswog2003,Mclaughlin2005,Surman2006,Surman2008,Metzger2008,Wanajo2012,Caballero2012}.
To date, only one neutrino-hydrodynamics calculation for merger remnants  has been published \citep{Dessart2009}.
This study was performed in two dimensions with the code VULCAN/2D and drew its initial conditions from 3D
SPH calculations with similar input physics, but without modelling the heating due to neutrinos \citep{Price2006}.
These calculations confirmed indeed that a neutrino-driven wind develops (with $\dot{M} \sim 10^{-3}$ \msun/s),
blown out into the funnel along the binary rotation axis that was previously thought to be practically baryon-free.
By baryon-loading the suspected launch path, this wind could potentially threaten the emergence of
the ultra-relativistic outflow that is needed for a short GRB. \cite{Dessart2009} therefore concluded that the launch
of a sGRB was unlikely to happen in the presence of the HMNS, but could possibly occur after the collapse to a black hole.\\
The aim of this study is to explore further neutrino-driven winds from compact binary mergers remnants.
We focus here on the phase where a HMNS is present in the centre and we assume
that it does not collapse during the time frame of our simulation, as in \cite{Dessart2009}. Given the
various stabilising mechanism such as thermal support, possibly magnetic fields and in particular
the strong differential rotation of the HMNS together with a lower limit on the maximum 
mass  in excess of $2.0$ \msun \citep{Demorest2010,Antoniadis2013}, we consider this as a very plausible 
assumption. We are mainly interested to see how robust the previous 2D results are with respect to a 
transition to three spatial dimensions. The questions about 
the understanding of the heavy element nucleosynthesis that occurs in compact binary mergers, 
the prediction of observable electromagnetic counterparts for the different outflows, 
and the emergence of sGRBs are the main drivers behind this work.\\

This paper is organized as follows.
In \refsec{sec: preliminary analysis and estimates},
we estimate the most important disc and wind time-scales. 
The details of our numerical model are explained in \refsec{sec: method}. In addition,
we briefly present the merger simulation, 
the outcome of which is used as initial condition for our study.
Our results are presented in \refsec{sec: results}. 
We briefly discuss in \refsec{sec:discussion} the nucleosynthesis in the neutrino-driven wind
and the properties of the radioactively powered, electromagnetic transients 
that result from them. Our major results are finally summarised in \refsec{sec:conclusions}.

%
%
\section{Analytical estimates}
\label{sec: preliminary analysis and estimates}

The properties of the remnant of a BNS merger can vary significantly 
\citep[see, for example, ][and references therein]{Rosswog2013,Bauswein2013a,Hotokezaka2013,Wanajo2014}, 
depending on the binary parameters (mass, mass ratio, eccentricity, spins etc.) and on the nuclear 
equation of state (hereafter, EoS). For our estimates and scaling relations, we use numerical values
that characterise our initial model, see \refsec{sec: initial conditions} for more details.\\
We consider a central HMNS of mass $M_{\rm ns} \approx 2.5$  \msun,
radius $R_{\rm ns} \approx 25 \, {\rm km}$  and temperature $k_{\rm B} T_{\rm ns} \approx 15 \, {\rm MeV} $. 
Inside of it, neutrinos are assumed to be in thermal equilibrium with matter. Under these conditions
the typical neutrino energy can be estimated as $E_{\nu,{\rm ns}} \sim \left( F_3(0)/F_2(0) \right) \, 
k_{\rm B}T_{\rm ns}\approx 3.15 \, k_{\rm B}T_{\rm ns} \approx 50 \, {\rm MeV}$,
where $F_n(0)$ is the Fermi integral of order $n$, evaluated for a vanishing degeneracy parameter.
The central object is surrounded by a geometrically thick disc of mass $M_{\rm disc} \approx 0.2$ \msun,
radius $R_{\rm disc} \approx 100 \, {\rm km}$ and height $H_{\rm disc} \approx 33 \, {\rm km}$.
The aspect ratio of the disc is then $H/R \approx 1/3$.
We assume a neutrino energy in the disc of $E_{\nu,{\rm disc}} \sim 15 \, {\rm MeV}$, 
comparable with the mean energy of the ultimately emitted neutrinos \citep[see, for example, ][]{Rosswog2013}.\\
Representative density values in the HMNS and in the disc are $\rho_{\rm ns} \sim 10^{14} {\rm g \, cm^{-3}}$ and 
$\rho_{\rm disc}\sim 5 \cdot 10^{11} {\rm g \, cm^{-3}}$, respectively.\\ 
The \emph{dynamical time-scale} $t_{\rm dyn}$ of the disc is set by the orbital Keplerian 
motion around the HMNS,
\be
 t_{\rm dyn} \sim \frac{2 \pi}{\Omega_K} \approx 0.011 \, {\rm s} \, 
 \left( \frac{M_{\rm ns}}{2.5 \ms} \right)^{-1/2} 
 \left( \frac{R_{\rm disc}}{100 \, {\rm km}} \right)^{3/2},
 \label{eqn: keplerian time-scale}
\ee
where $\Omega_K$ is the Keplerian angular velocity.\\
On a time-scale longer than $t_{\rm dyn}$, viscosity drives radial motion.
Assuming it can be described by an $\alpha$-parameter model \citep{Shakura1973},
we estimate the \emph{lifetime of the accretion disc} $t_{\rm disc}$ as 
\begin{eqnarray}
 \lefteqn{t_{\rm disc} \sim \alpha^{-1} \left( \frac{H}{R} \right)^{-2} \Omega_{K}^{-1} \approx 
 0.3 \, {\rm s} \, \left( \frac{\alpha}{0.05} \right)^{-1} \left( \frac{H/R}{1/3} \right)^{-2} } \nonumber \\
 && \left( \frac{M_{\rm ns}}{2.5 \Ms} \right)^{-1/2} \left( \frac{R_{\rm disc}}{100 \, {\rm km}} \right)^{3/2}.
 \label{eqn: viscosity time}
\end{eqnarray}
The \emph{accretion rate} on the HMNS $\dot{M}$ is then of order
\begin{eqnarray}
 \lefteqn{\dot{M} \sim \frac{M_{\rm disc}}{t_{\rm disc}} \approx 0.64 \, \frac{\ms}{\rm s} \, 
 \left( \frac{M_{\rm disc}}{0.2 \, \Ms} \right) \left( \frac{\alpha}{0.05} \right) } \nonumber \\
 && \left( \frac{H/R}{1/3} \right)^{2} \left( \frac{M_{\rm ns}}{2.5 \ms} \right)^{1/2} 
 \left( \frac{R_{\rm disc}}{100 \, {\rm km}} \right)^{-3/2}.
 \label{eqn: mdot}
\end{eqnarray}

Neutrinos are the major cooling agent of the remnant.
Neutrino scattering off nucleons is one of the major sources of {\em opacity} for all neutrino species\footnote{In the
case of $\nue$'s, the opacity related with absorption by neutrons is even larger. Nevertheless, it is still comparable 
to the scattering off nucleons.}
and the corresponding mean free path can be estimated as
\be
 \lambda_{N \nu} 
 \approx 7.44 \cdot 10^3 \, {\rm cm} \, \left( \frac{\rho}{10^{14} \, {\rm g/cm^3}} \right)^{-1} 
 \left( \frac{E_{\nu}}{10 \, {\rm MeV}} \right)^{-2},
 \label{eqn: lambda nu}
 \ee
where $\rho$ is the matter density and $E_{\nu}$ is the typical neutrino energy.
The large variation in density between the HMNS and the disc
suggests to treat these two regions separately.\\
For the central compact object, the \emph{cooling time-scale} $t_{\rm cool,ns}$ 
is governed by neutrino diffusion (see, for example, \cite{Rosswog2003}).
If $\tau_{\nu,{\rm ns}}$ is the neutrino \emph{optical depth} inside the HMNS, then
\be
t_{\rm cool, ns} \sim 3 \, \frac{\tau_{\nu,{\rm ns}} \, R_{\rm ns}}{c}.
\label{eqn: diffusion time-scale estimate}
\ee
If we assume $\tau_{\nu,{\rm ns}} \sim R_{\rm ns}/ \lambda_{N \nu}$,
\be
 t_{\rm cool,ns} \sim 1.88 \, {\rm s} \left( \frac{R_{\rm ns}}{25 \, {\rm km}} \right)^2
 \left(\frac{\rho_{\rm ns}}{10^{14}{\rm g/cm^3}} \right) 
 \left(\frac{k_{\rm B}T_{\rm ns}}{15 \, {\rm MeV}} \right)^2.
 \label{eqn: diffusion time HMNS}
\ee
The neutrino luminosity coming from the HMNS is powered by an internal energy reservoir $\Delta E_{\rm ns}$.
We estimate it as the difference between the internal energy of a hot and of a cold HMNS.
For the first one, we consider typical profiles of a HMNS obtained from a BNS merger simulation.
For the second one, we set $T = 0$ everywhere inside it.
Under these assumptions, $\Delta E_{\rm ns} \approx 0.30 \, E_{\rm int, HMNS} \approx 3.4 \cdot 10^{52} {\rm erg} $,
and the associated HMNS neutrino luminosity (integrated over all neutrino species) is approximately
\begin{eqnarray}
 \lefteqn{L_{\nu,{\rm ns}} \sim \frac{\Delta E_{\rm ns}}{t_{\rm diff,ns}} \approx 
 1.86 \cdot 10^{52}\,\frac{\rm erg}{\rm s} \left( \frac{\Delta E_{\rm ns}}{3.5 \cdot 10^{52} {\rm erg} } \right)} \nonumber \\ 
 && \left( \frac{R_{\rm ns}}{25 \, {\rm km}} \right)^{-2}
    \left(\frac{\rho_{\rm ns}}{10^{14}{\rm g/cm^3}} \right)^{-1}
    \left(\frac{k_{\rm B}T_{\rm ns}}{15 \, {\rm MeV}} \right)^{-2}.
 \label{eqn: HMNS luminosity}                           
\end{eqnarray}
The disc diffusion time-scale can be estimated using an analogous to 
\refeq{eqn: diffusion time-scale estimate}:
\begin{eqnarray}
 \lefteqn{t_{\rm cool,disc} \sim 3 \, \frac{\tau_{\nu,{\rm disc}} \, H_{\rm disc}}{c} \approx 1.68 \, {\rm ms} \,
                       \left( \frac{H_{\rm disc}}{33 \, {\rm km}} \right)^2} \nonumber \\
 && \left(\frac{\rho_{\rm disc}}{5 \cdot 10^{11}{\rm g/cm^3}} \right)
    \left(\frac{E_{\nu,{\rm disc}}}{15 \, {\rm MeV}} \right)^2.
\label{eqn: diffusion time disc}
\end{eqnarray}
Due to this fast cooling time-scale, a persistent neutrino luminosity from the disc requires a constant 
supply of internal energy. In an accretion disc, this is provided by the \emph{accretion mechanism}: 
while matter falls into deeper Keplerian orbits, the released 
gravitational energy is partially ($\sim 50$ per cent) converted into
internal energy. 
If $R_{\rm disc} \sim 100 \, {\rm km}$ denotes the typical initial distance inside the disc, and
the radius of the HMNS is assumed to be the final one, then
$\Delta E_{\rm grav} \sim \left( G M_{\rm ns} M_{\rm disc} / R_{\rm ns}\right)$,
where we have used $R_{\rm ns}^{-1} \gg R_{\rm disc}^{-1}$.
The neutrino luminosity for the accretion process is approximately
\begin{eqnarray}
 \lefteqn{L_{\nu,{\rm disc}} \sim 0.5 \, \frac{\Delta E_{\rm grav}}{t_{\rm disc}} \approx 8.35 \cdot 10^{52} \, \frac{\rm erg}{\rm s}
     \left( \frac{M_{\rm ns}}{2.5 \, \Ms}  \right)^{3/2}
     \, \left( \frac{\alpha}{0.05 }  \right)} \nonumber \\
  && \left( \frac{M_{\rm disc}}{0.2 \, \Ms}  \right)
     \left( \frac{H/R}{1/3}  \right)^2
     \left( \frac{R_{\rm disc}}{100 \, {\rm km}}  \right)^{-3/2}
     \left( \frac{R_{\rm ns}}{25 \, {\rm km}}  \right)^{-1}.
\label{eqn: disc luminosity}
\end{eqnarray}\\

Note that during the disc accretion phase $L_{\nu,{\rm disc}}$ is  larger 
than  $L_{\nu,{\rm ns}}$. Together, the HMNS and the disc release neutrinos at
a luminosity of $\sim 10^{53}$ erg/s, consistent with the simple estimate from the introduction.\\
Due to the density (opacity) structure of the disc, the neutrino emission is expected to be anisotropic, with a larger  
luminosity in the polar directions ($\theta = 0$ and $\theta = \pi$), compared to the one 
along the equator ($\theta = \pi/2$), see also \cite{Rosswog2003a,Dessart2009}. For a simple
model of this effect, we assume that the disc creates an axisymmetric shadow area 
across the equator, while the emission is uniform outside this area. 
The amplitude of the shadow is $2 \, \theta_{\rm disc}$, where
$\tan{\theta_{\rm disc}} = (H/R)$. 
Then, we define an isotropised axisymmetric luminosity $L_{\nu,{\rm iso}}(\theta)$ as 
(see the sketch on the right in \reffig{fig: cartoon}):
 \be
  L_{\nu,{\rm iso}}(\theta) = 
  \left\{
  \begin{array}{rl} 
   \xi \, L_{\nu} & \quad \mbox{for } \left| \theta - \pi/2 \right| > \theta_{\rm disc}   \\
   0              & \quad \mbox{for } \left| \theta - \pi/2 \right| \leq \theta_{\rm disc}.
  \end{array}
  \right.
 \ee
The value of $\xi$ is set by the  normalisation of $L_{\nu,{\rm iso}}$ over 
the whole solid angle $\Omega$, $\int_{\Omega} L_{\nu,{\rm iso}} \, {\rm d}\Omega = L_{\nu}$:
\be
 \xi = \frac{1}{1-\sin{\theta_{\rm disc}}}.
\ee
For $(H/R) \approx 1/3$, one finds $\theta_{\rm disc} \approx \pi/10$ and $\xi \approx 1.5$.

After having determined approximate expressions for the neutrino luminosities, 
we are ready to estimate
the relevant time-scale for the formation of the $\nu$-driven wind.\\
We define the \emph{wind time-scale} $t_{\rm wind}$ 
as the time necessary for the matter 
to absorb enough energy 
to overcome the gravitational well generated by the HMNS. 
This energy deposition happens inside the disc and it is due to 
the re-absorption of neutrinos emitted
at their last interaction surface.
Thus,
\be
 t_{\rm wind} \sim e_{\rm grav}/\dot{e}_{\rm heat}, 
\label{eqn: wind time-scale}
\ee
where $e_{\rm grav} \approx G M_{\rm ns}/R$ is the specific gravitational energy,
and $\dot{e}_{\rm heat}$ is the specific heating rate provided by neutrino absorption
at a radial distance $R$ from the centre:
\be
\dot{e}_{\rm heat} \sim k \,\frac{L_{\nue,{\rm iso}}(\left| \theta - \pi/2 \right| > \theta_{\rm disc} )}
                     {4 \, \pi \, R^2} .
\label{eqn: heating rate}
\ee
In the equation above we have assumed that
$L_{\nue} \approx L_{\nueb} \sim \left( L_{\nu,{\rm ns}} + L_{\nu,{\rm disc}} \right)/3 $.
If $k \approx 5.65 \cdot 10^{-20} \, {\rm cm^{2}} \, {\rm g^{-1} \, MeV^{-2}} \, E_{\nu}^2 \, $
is the typical absorptivity on nucleons \citep{Bruenn1985}, 
the heating rate can be re-expressed as
\begin{eqnarray}
\lefteqn{\dot{e}_{\rm heat} \sim 4.6 \cdot 10^{20} \frac{\rm erg}{\rm g \cdot s} \left( \frac{R}{100 \, {\rm km}} \right)^{-2}} \nonumber \\ 
  && 
     \left( \frac{L_{\nue}}{3 \cdot 10^{52} \, {\rm erg/s}} \right)
     \left( \frac{\xi}{1.5} \right)
     \left( \frac{E_{\nu,{\rm disc}}}{15 \, {\rm MeV}} \right)^{2}.
 \label{eqn: heating rate estimation}
\end{eqnarray}
Finally, the wind time-scale, \refeq{eqn: wind time-scale}, becomes
\begin{eqnarray}
 \lefteqn{t_{\rm wind} \sim 0.07 \, {\rm s} \, \left( \frac{M_{\rm ns}}{2.5 \, \Ms}    \right)
                                     \left( \frac{R}{100 \, {\rm km}} \right)} \nonumber \\
   && \left( \frac{L_{\nue}}{3 \cdot 10^{52} \, {\rm erg/s}} \right)^{-1}
      \left( \frac{\xi}{1.5} \right)^{-1}
      \left( \frac{E_{\nu,{\rm disc}}}{15 \, {\rm MeV}} \right)^{-2}.
\end{eqnarray}
Since $t_{\rm wind} < t_{\rm disc}$, neutrino heating  
can drive a wind within the lifetime of the disc. 
Moreover, since the disc provides a substantial fraction of the total neutrino luminosity,
a wind can form also in the absence of the HMNS. 

Of course, the neutrino emission processes are much more complicated than what can be captured by 
these simple estimates. Nevertheless, they provide a reasonable first guidance for the qualitative understanding
of the remnant evolution.

%
%

\section{Numerical model for the remnant evolution}
\label{sec: method}

%
%
\subsection{Hydrodynamics}
\label{sec:hydro}

We perform our simulations with the \texttt{FISH} code \citep{Kaeppeli2011}.
\texttt{FISH} is a parallel grid code that solves the equations of ideal, Newtonian 
hydrodynamics (HD) \footnote{\texttt{FISH} can actually solve the
equations of ideal magnetohydrodynamics. However, we have not included
magnetic fields in our current setup.}:
\be
\label{eq:hd_0010}
  \frac{\partial \rho}{\partial t}
    + \nabla \cdot \left( \rho {\bf v} \right)
   =  0 
\ee
\be
   \label{eq:hd_0020}
  \frac{\partial \rho {\bf v}}{\partial t}
    + \nabla \cdot \left( \rho {\bf v} \otimes {\bf v} \right)
    + \nabla p
   =  - \rho \nabla \phi + \rho \, \left( \frac{{\rm d}{\bf v}}{{\rm d}t} \right)_{\nu} 
\ee
\be
\label{eq:hd_0030}
  \frac{\partial E}{\partial t}
    + \nabla \cdot \left[ \left( E + p \right) {\bf v} \right]
   =  - \rho {\bf v} \nabla \phi
        + \rho \, \left( \frac{{\rm d}e}{{\rm d}t} \right)_{\nu} 
\ee
\be
\label{eq:hd_0040}
  \frac{\partial \rho Y_e}{\partial t}
    + \nabla \cdot \left( \rho Y_e {\bf v} \right)
   =  \rho \, \left( \frac{{\rm d} Y_e}{{\rm d}t} \right)_{\nu}
\ee
Here $\rho$ is the mass density, ${\bf v}$ the velocity,
$E = \rho e + \rho v^2/2$ the total energy density (i.e., the sum of
internal and kinetic energy density), $e$ the specific internal energy, 
$p$ the matter pressure and $Y_e$ the electron fraction.
The code solves the HD equations with a second-order accurate finite
volume scheme on a uniform Cartesian grid.
The source terms on the right hand side stem from gravity and from neutrino-matter interactions.
We notice that the viscosity of our code is of numerical nature, 
while no physical viscosity is explicitly included.
The neutrino source terms will be discussed in detail in \refsec{sec:neutrino treatment}.
The gravitational potential $\phi$ obeys the Poisson equation
\be
\label{eq:hd_0050}
  \nabla^2 \phi = 4 \pi G \rho
  ,
\ee
where $G$ is the gravitational constant.
The merger of two neutron stars with equal masses is expected 
to form a highly axisymmetric remnant.
We exploit this approximate invariance by solving the Poisson equation in cylindrical symmetry.
This approximation results in a high gain in computational efficiency, given the
elliptic (and hence global) nature of \refeq{eq:hd_0050}.
To this end, we \emph{conservatively} average the three-dimensional density
distribution onto an axisymmetric grid, having the HMNS rotational axis as the
symmetry axis.
The Poisson equation is then solved with a fast multigrid
algorithm \citep{Press1992}, and
the resulting potential is interpolated back on the
three-dimensional grid.

The HD equations are closed by an
EoS relating the internal energy to the pressure.
In our model, we use the TM1 EoS description of nuclear matter supplemented with
electron-positron and photon contributions, in tabulated form
\citep{Timmes2000,Hempel2012}. This description is equivalent to one provided by the 
Shen et al. EoS \citep{Shen1998a,Shen1998b} in the high density part.

%
%
\subsection{Neutrino treatment}
\label{sec:neutrino treatment}

In general, the multi-dimensional neutrino transport is described 
by the equation of radiative transfer \citep[see, for example, ][]{Mihalas1984}.
Instead of a direct solution of this equation,
which is computationally very expensive in large multi-dimensional simulations, we employ a relatively inexpensive,
effective neutrino treatment. Our goal is to provide expressions for the neutrino source terms, 
assuming to know qualitatively the solution of the radiative transfer equation 
in different parts of the domain.
Our treatment is a spectral extension of previous grey leakage schemes 
\citep{Ruffert1996,Rosswog2003}. However, differently from its predecessors, 
it includes also {\em spectral absorption terms in the optically thin regime}.
The treatment has been developed and tested against detailed Boltzmann neutrino transport 
for spherically symmetric core collapse supernova models.
For two tested progenitors ($15$ \msun~ and $40$ \msun~ zero age main sequence stars),
the neutrino luminosities and the shock positions agree within 20 per cent with the corresponding values
obtained by Boltzmann transport, for a few hundreds of milliseconds after core bounce.
A detailed description with
tests will be discussed in a separate paper (Perego et al. 2014, in preparation).
Here we provide a summary of the method and we refer to it as 
an Advance Spectral Leakage (ASL) scheme\footnote{The ASL scheme allows also the modelling
of the neutrino trapped component. However, since this component was not included
in the study of the merger process that provided our initial conditions, we neglect it here.}.

The neutrino energy is discredited in 12 geometrically increasing energy bins, chosen in the range 
$2 \, {\rm MeV} \leq E_{\nu} \leq 200 \, {\rm MeV}$. 
The ASL scheme includes the reactions listed in \reftab{table: nu reactions}. 
They correspond to the reactions that we expect to be more relevant in hot and dense matter.
Neutrino pair annihilation is included only as a source of opacity in optically thick conditions.
Due to the geometry of the emission, it is also supposed to be 
important in optically thin conditions (see, for example, \cite{Janka1991,Burrows2006}. For 
the application to the BNS merger scenario, see \cite{Dessart2009} and references therein). 
Therefore, our numbers concerning the mass loss
$\dot{M}$ need to be considered as lower limits on the true value.
The full inclusion of this process in our model will be performed in a future step. 
\begin{table}
 \begin{center}
    \begin{tabular}{| l | c | l |}
    \hline
    Reaction                                   & Roles   & Ref. \\ \hline
    $e^- + p \leftrightarrow n + \nue $            & O,T,P   & a    \\    
    $e^+ + n \leftrightarrow p + \nueb $           & O,T,P   & a    \\         
    $e^- + (A,Z) \leftrightarrow \nue + (A,Z-1)$   & T,P     & a    \\         
    $N + \nu \leftrightarrow N + \nu $             & O       & a    \\         
    $(A,Z) + \nu \leftrightarrow (A,Z) + \nu $     & O       & a    \\ 
    $e^+ + e^- \leftrightarrow \nu + \nub $        & T,P     & a,b  \\ 
    $N + N \leftrightarrow N + N + \nu + \nub $    & T,P     & c    \\ \hline
    \end{tabular}
  \end{center}
\caption{List of the neutrino reactions included in the simulation (left column; 
   $\nu \equiv \nue,\nueb,\numt$), of their major effects
   (central column; O stands for opacity, P for neutrino
   production, T for neutrino thermalisation), and of the references for the implementation (right column):
   ``a'' corresponds to \protect\cite{Bruenn1985}, ``b'' to \protect\cite{mezzacappa1993}, 
   and ``c'' to \protect\cite{Hannestad1998}.
   }
  \label{table: nu reactions}
\end{table}

The ASL scheme models explicitly three different neutrino species: $\nue$, $\nueb$, and $\numt$.
The species $\numt$ is a collective species for $\mu$ and $\tau$ (anti-)neutrinos, 
that contributes only as a source of cooling in the energy equation.
As a consequence of the distinction between emission and absorption processes, and between different
neutrino species, the source terms in \refeq{eq:hd_0020}-\refeq{eq:hd_0040} 
can be split into different contributions. For the electron fraction,
\be
 \left( \frac{{\rm d}Y_e}{{\rm d}t} \right)_\nu = - m_b \, \left[ \left( R^0_{\nue} - R^0_{\nueb} \right) + \left( H^0_{\nue} - H^0_{\nueb} \right) \right],
 \label{eqn: yedot contributions}
\ee
where $R^0_{\nu}$ and $H^0_{\nu}$ denote the specific particle emission and absorption rates 
for a neutrino type $\nu$ respectively, and $m_b$ is the baryon mass (with $m_b c^2 = 939.021 \, {\rm MeV} $).
For the specific internal energy of the fluid,
\be
 \left( \frac{{\rm d}e}{{\rm d} t} \right)_\nu =   - \left( R^1_{\nue} + R^1_{\nueb} + 4 \, R^1_{\numt}  \right) + H^1_{\nue} + H^1_{\nueb},
 \label{eqn: edot contributions}
\ee
where $R^1_{\nu}$ and $H^1_{\nu}$ indicate the specific energy emission and absorption rates, respectively.
The factor 4 in front of $R^1_{\numt}$ accounts for the four different species modelled 
collectively as $\numt$.  
And, finally, for the fluid velocity,
\be
 \left( \frac{{\rm d}\mathbf{v}}{{\rm d}t} \right)_{\nu} = 
 \left( \frac{{\rm d}\mathbf{v}}{{\rm d}t} \right)_{\nue} + 
 \left( \frac{{\rm d}\mathbf{v}}{{\rm d}t} \right)_{\nueb}.
 \label{eqn: vdot contributions}
\ee
is the acceleration provided by the momentum transferred by 
the absorption of $\nue$'s and $\nueb$'s in the optically thin region.
Since the  trapped neutrino component is not dynamically modelled, we neglect
the related neutrino stress in optically thick conditions. As a consequence,
$\numt$'s do not contribute to the acceleration term.\\
For each neutrino $\nu$ species, the  \emph{luminosity} ($L_{\nu}$) and 
\emph{number luminosity} ($L_{N,\nu}$) 
are calculated as:
\be
 L_{\nu} =  \int_V \: \rho \left( R^1_{\nu} - H^1_{\nu} \right) \, {\rm d}V 
 \label{eqn: nu luminosity}
\ee
and
\be
 L_{N,\nu} =  \int_V \: \rho \left( R^0_{\nu} - H^0_{\nu} \right)  \, {\rm d}V. 
 \label{eqn: nu N luminosity}   
\ee
where $V$ is the volume of the domain.
The explicit distinction between the emission and the absorption contributions, as well as 
their dependence on the spatial position, allows the introduction of two supplementary 
luminosities:\\
1) The \emph{cooling luminosities}, $L_{\nu,{\rm cool}}$ and $L_{N,\nu,{\rm cool}}$, 
obtained by neglecting the heating rates $H_{\nu}^1$ and $H_{\nu}^0$ in \refeq{eqn: nu luminosity} 
and \refeq{eqn: nu N luminosity}, respectively.\\
2) The \emph{HMNS luminosities}, $L_{\nu,{\rm HMNS}}$ and $L_{N,\nu,{\rm HMNS}}$, obtained by 
restricting the volume integral in  \refeq{eqn: nu luminosity} and \refeq{eqn: nu N luminosity}
to $V_{\rm HMNS}$, the volume of the central object.
Due to the continuous transition between the HMNS and the disc, the definition of $V_{\rm HMNS}$ 
is somewhat arbitrary. We decide to include also
the innermost part of the disc, delimited by a density contour of $5 \times 10^{11} {\rm g \, cm^{-3}}$.
This corresponds to the characteristic density close to the innermost stable orbit for a torus 
accreting on stellar black holes. 
It is also comparable with the surface density of a cooling proto-neutron star.
For each luminosity we associate a \emph{neutrino mean energy}, 
defined as $\langle E_{\nu} \rangle \equiv L_{\nu}/L_{N,\nu}$.\\
Since the scheme is spectral, all the terms on the right hand side of \refeq{eqn: yedot contributions}, 
\refeq{eqn: edot contributions} and \refeq{eqn: vdot contributions}
are energy-integrated values of spectral emission ($r_{\nu}$), absorption ($h_{\nu}$) and stress ($\mathbf{a}_{\nu}$) rates:
\be
R^n_{\nu} =  \int_0^{+\infty} \! {r_{{\nu}}}\, E^{n+2} \, dE , \label{eqn: particle cooling rate}
\ee
\be
H^n_{\nu} =  \int_0^{+\infty} \! {h_{{\nu}}}\, E^{n+2} \, dE , \label{eqn: particle heating rate}
\ee
\be
\left( \frac{{\rm d} \mathbf{v}}{{\rm d} t} \right) _{\nu} = \int_{0}^{+\infty} \mathbf{a}_{\nu} \, E^2 \; {\rm d} E. \label{eqn: stress rate}
\ee
The calculation of $r_{\nu}$, $h_{\nu}$ and $\mathbf{a}_{\nu}$ is the ultimate purpose of the ASL scheme.

 \begin{figure}
 \begin{center}
 \includegraphics[width = 0.8 \linewidth]{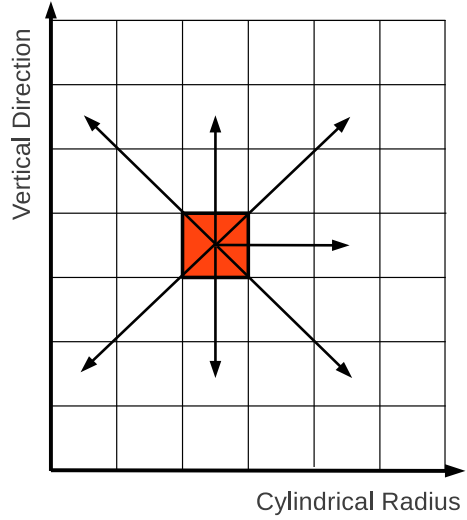}
 \end{center}
 \caption{Schematic plot of the seven directions (paths) used to compute the optical depth at each point
 of the cylindrical domain.}
 \label{fig: tau_paths}
 \end{figure}
 
The neutrino optical depths $\tau_\nu$'s play a central role in our scheme.
We distinguish between the {\em scattering} ($\tau_{\nu,{\rm sc}}$) 
and the {\em energy} ($\tau_{\nu,{\rm en}}$) spectral optical depth.
The first one is obtained by summing all the relevant neutrino processes:
\be
\label{eqn: tau cl def}
{\rm d} \tau_{\nu,{\rm sc}} = \rho \left( k_{\rm sc} + k_{\rm ab} \right)  \, {\rm d}s
\ee
where ${\rm d}s$ is an infinitesimal line element, and $k_{\rm ab}$ and $k_{\rm sc}$ are the neutrino opacities 
for absorption and scattering, respectively.
For the second, more emphasis is put on those inelastic processes, that are effective in keeping 
neutrinos in thermal equilibrium with matter.
In this case, we have 
\be
\label{eqn: tau eff def}
{\rm d} \tau_{\nu,{\rm en}} = \rho \sqrt{ k_{\rm ab} \left(  k_{\rm sc} + k_{\rm ab}  \right)} \, {\rm d} s,
\ee
where we have considered absorption processes as inelastic, and scattering processes as elastic\footnote{This is not true in
general. However, it applies to the set of reactions we have chosen for our model. See \reftab{table: nu reactions}.}.
The values of the spectral $\tau_{\nu}$'s at each point are calculated using a local ray-by-ray method. 
It consists of integrating \refeq{eqn: tau cl def} and 
\refeq{eqn: tau eff def} 
along several predefined paths and taking the minimum values among them.
These paths are straight oriented segments, connecting the considered point with 
the edge of the computational domain.
Due to the intrinsically global character of
these integrations, we decided to exploit also here the expected symmetry of
the remnant, and to calculate $\tau_{\nu}$ in axial symmetry.
The seven different paths we explore in the $(R_{\rm cyl}-z)$ plane are shown in \reffig{fig: tau_paths}.
As a future step, we plan to include the more sophisticated and geometrically more flexible
MODA methods \citep[][]{Perego2014} to compute $\tau_{\nu}$.\\
The optical depths vary largely and they decrease, following the density profile, 
proceeding from the HMNS to the edge of the remnant. To characterise this behaviour, we define 
the unit vector
\be
\mathbf{\hat{n}}_{\tau} \equiv - \nabla \tau_{\nu,{\rm sc}} / \left| \nabla \tau_{\nu,{\rm sc}} \right|,
\ee
computed at each point of the domain from finite differences on the grid. 
This vector will be crucial later to model the diffusion and the final
emission of the neutrinos.\\
The surfaces where $\tau_{\nu}$ equals 2/3 are defined as {\it neutrino surfaces}.
The neutrino surfaces obtained from 
$\tau_{\nu,{\rm sc}}$ can be understood as the \emph{last scattering surfaces}; the ones derived 
from $\tau_{\nu,{\rm en}}$ correspond to the surfaces where neutrinos decouple thermally from matter,
and they are often called \emph{energy surfaces} \citep[see, for example,][]{Raffelt2001}.
According to the value of $\tau_{\nu,{\rm sc}}$, we distinguish between three disjoint volumes:
1) $V_{\rm thin}$, for the optically thin region ($\tau_{\nu,{\rm sc}} \ll 2/3$); 2) $V_{\rm surf}$, for the neutrino 
surface region\footnote{In principle, the neutrino surfaces should have no volume. However, due
to the discretisation on the (axisymmetric) grid we adopted to calculate $\tau$, 
every neutrino surface is replaced by a shell of width 
$\sim \Delta x$. This thin layer is formed by the cells $\mathbf{x}$ 
inside which $\tau$ is expected to become equal to 2/3.} ($\tau_{\nu,{\rm sc}} \sim 2/3$);
3) $V_{\rm thick}$, for the optically thick region ($\tau_{\nu,{\rm sc}} \gg 2/3$). 
Obviously, $V = V_{\rm thick} \, \cup \, V_{\rm surf} \, \cup \, V_{\rm thin} $.\\

\begin{table}
 \begin{center}
    \begin{tabular}{| l | l | l |}
    \hline
    Quantity                      & Definition                & Related quantities \\ \hline
    $j_{\rm em}$                  & emissivity                & $r_{\nu,{\rm prod}}$    \\ 
    $k_{\rm ab}$                  & absorption opacity        & $\lambda_{\nu}$, $\tau_{\nu}$, $h_{\nu}$, $\mathbf{a}_{\nu}$    \\ 
    $k_{\rm sc}$                  & scattering opacity        & $\lambda_{\nu}$, $\tau_{\nu}$    \\ \hline        
    $\lambda_{\nu}$               & mean free path            & $r_{\nu,{\rm diff}}$    \\
    $\tau_{\nu}$                  & optical depth             & $\hat{\mathbf{n}}_{\tau}$, $\hat{\mathbf{n}}_{\rm path}$, $r_{\nu,{\rm diff}}$, $r_{\nu,{\rm ult}}$, \\
                                  &                           & $\quad h_{\nu} $, $\mathbf{a}_{\nu} $    \\ 
    $\hat{\mathbf{n}}_{\tau}$     & opposite $\tau$ gradient  & $\hat{\mathbf{n}}_{\rm path}$, $n_{\nu} $, $\mathbf{s}_{\nu} $    \\ 
    $\hat{\mathbf{n}}_{\rm path}$ & diffusion direction       & $r_{\nu,{\rm ult}}$  \\ \hline
    $r_{\nu,{\rm prod}}$          & production rate           & $r_{\nu}$    \\ 
    $r_{\nu,{\rm diff}}$          & diffusion rate            & $r_{\nu}$    \\ 
    $r_{\nu}$                     & emission rates            & $r_{\nu,{\rm ult}}$, $( {\rm d} Y_e/{\rm d}t)_{\nu}$, $ ({\rm d}e/{\rm d}t)_{\nu}$  \\ \hline
    $r_{\nu,{\rm ult}} $          & ultimate emission rates   & $n_{\nu}$, $\mathbf{s}_{\nu} $   \\
    $n_{\nu} $                    & particle density          & $h_{\nu} $ \\
    $\mathbf{s}_{\nu} $           & momentum density          & $\mathbf{a}_{\nu}$ \\
    $h_{\nu} $                    & absorption rate           & $( {\rm d}Y_e/{\rm d}t)_{\nu}$, $ ({\rm d}e/{\rm d}t)_{\nu}$ \\
    $\mathbf{a}_{\nu} $           & stress                    & $( {\rm d} \mathbf{v}/ {\rm d}t)_{\nu}$ \\ \hline
    \end{tabular}
  \end{center}
\caption{List of the most important spectral quantities appearing in the ASL scheme (left column) and
their definition (central column). In the right column, we list the relevant
quantities (spectral quantities and source terms) that depend directly on each table entry.
See the text for more details.}
  \label{table: ASL summary}
\end{table}

After having introduced $\tau_\nu$, we can now explain in which way the neutrino rates
are calculated within the ASL scheme.
In \reftab{table: ASL summary} we have summarised the most important quantities, their
definitions and relations in the context of the ASL scheme.\\
The spectral emission rates 
$r_{\nu}$ are calculated as smooth interpolation between diffusion ($r_{\nu,{\rm diff}}$) 
and production ($r_{\nu,{\rm prod}}$) spectral rates: the first ones are the relevant rates 
in the optically thick regime, the latter in the optically transparent region.\\
We compute $r_{\nu,{\rm prod}}$ and $r_{\nu,{\rm diff}}$ as
\begin{eqnarray}
\lefteqn{r_{\nu,{\rm prod}} = \frac{4 \pi}{\left( hc \right)^3} \frac{ j_{\rm em} }{ \rho },}
\label{eqn: production rates} \\
\lefteqn{r_{\nu,{\rm diff}} =  \frac{4 \pi}{ \left( hc \right)^3} \frac{ f_{\nu}^{\rm FD}}{ \rho \, t_{\nu,{\rm diff}}}.}
\label{eqn: diffusion rates}
\end{eqnarray}
$j_{\rm em}$ is the neutrino spectral emissivity, while 
$f_{\nu}^{\rm FD}$ is the Fermi-Dirac distribution function 
for a neutrino gas in thermal and weak equilibrium with matter. 
$t_{\nu,{\rm diff}}$ is the local diffusion time-scale, calculated as
\be
 t_{\nu,{\rm diff}} = \alpha_{\rm diff} \frac{\tau_{\nu,{\rm sc}}^2 \, \lambda_{\nu,{\rm sc}}}{c}
 \label{eqn: diffusion timescale}
 \ee
where $\lambda_{\nu,{\rm sc}} = \left( \rho \left( k_{\nu,{\rm ab}} + k_{\nu,{\rm sc}} \right) \right)^{-1}$ is the 
total mean free path. 
$\alpha_{\rm diff}$ is a constant set to 3.
The interpolation formula for $r_{\nu}$ is provided by half of 
the harmonic mean between the production and diffusion rates.\\
We compute the spectral heating rate as the properly normalised product of the 
absorption opacity $k_{\nu,{\rm ab}}$ and of the spectral neutrino density $n_{\nu}$:
\begin{eqnarray}
 h_{\nu} = c\,  k_{\nu,{\rm ab}} \, n_{\nu} \, \mathcal{F}_{e,\nu} \, \mathcal{H}.
 \label{eqn: heating term outside}
\end{eqnarray}
$\mathcal{H} \equiv \exp(- \tau_{{\nu},\rm sc})$ is an exponential cut off that ensures 
the application of the heating term only outside the neutrino surface, and
$\mathcal{F}_{e,\nu}$ is the Pauli blocking factor for electrons or positrons in the final state. 
$n_{\nu}$ is defined so that the energy-integrated particle density $N_{\nu}$ is given by:
\be
 \label{eqn: spectral neutrino density}
 N_{\nu} = \int_{0}^{+ \infty}  \, n_{\nu} \, E^2 \, {\rm d} E.
\ee
The stress term is calculated similarly to the neutrino heating rate:
\be
 \mathbf{a}_{\nu} = c \, k_{\nu,{\rm ab}} \, \mathbf{s}_{\nu} \, \mathcal{F}_{e,\nu} \; \mathcal{H},
 \label{eqn: spectral stress rate}
\ee
where $\mathbf{s}_{\nu}$ is the spectral density of linear momentum associated with the streaming neutrinos,
while $\mathcal{H}$ and $\mathcal{F}_{e,\nu}$ are defined as in \refeq{eqn: heating term outside}.\\

\begin{figure}
\begin{center}
\includegraphics[width = \linewidth]{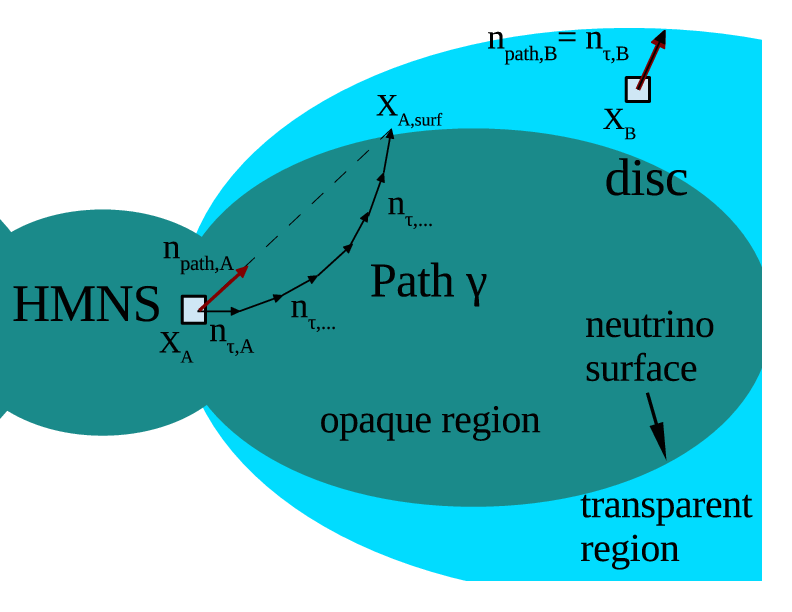}
\end{center}
\caption{Schematic representation of the procedure to calculate the ultimate emission rates at the 
neutrino surface and in the optically thin region, $r_{\nu,{\rm ult}}$, from the emission
rates, $r_{\nu}$. The thin black arrows represent the inverse of the gradient of 
$\tau_{\rm sc}$ ($\mathbf{\hat{n}}_{\tau}$),
while the thick red arrow is $\mathbf{\hat{n}}_{\rm path}$.
Label $x_{A}$ refers to a point inside the neutrino surface (opaque region),
while $x_{B}$ is a point inside the disc, but in the optically thin zone, for which
$\mathbf{\hat{n}}_{\rm tau} = \mathbf{\hat{n}}_{\rm path}$. See the text for more details.}
\label{fig: path}
\end{figure}

The quantities $n_{\nu}$ and $\mathbf{s}_{\nu}$ are computed using a multidimensional 
ray-tracing algorithm.
This algorithm assumes that neutrinos (possibly, after having diffused from the optically thick region) 
are \emph{ultimately} emitted isotropically at the neutrino surface and in the
optically transparent region.
If we define $l_{\nu}({\mathbf{x}',\mathbf{\hat{n}}})$ 
as the specific rate per unit solid angle of the radiation 
emitted from a point $\mathbf{x}' \in \left( V_{\rm surf} \cup V_{\rm thin} \right)$, in the direction $\mathbf{\hat{n}}$, then 
\be
\label{eqn: nu_dens_eq}
 n_{\nu}(\spc) = \int_{V_{\rm surf} \, \cup \, V_{\rm thin}} \:
 \rho \, \frac{ l_{\nu} \left(\mathbf{x}',\mathbf{\hat{n}(\mathbf{x},\mathbf{x}')} \right)}{c \left| \mathbf{x}' - \mathbf{x} \right|^2} {\rm d}^3 \mathbf{x}'
\ee
and
\be
 \mathbf{s}_{\nu}(\spc) = \int_{V_{\rm surf} \, \cup \, V_{\rm thin}} \:
 \rho \,  \frac{l_{\nu} 
 \left( \mathbf{x}',\mathbf{\hat{n}(\mathbf{x},\mathbf{x}')} \right)}{c \left| \mathbf{x}' - \mathbf{x} \right|^2}
 \, \frac{E}{c} \, \mathbf{\hat{n}(\mathbf{x},\mathbf{x}')} \; {\rm d}^3 \mathbf{x}'. \label{eqn: spectral stress vector}
\ee
where $\mathbf{\hat{n}}(\mathbf{x},\mathbf{x}') = 
(\mathbf{x}' - \mathbf{x})/( \left| \mathbf{x}' - \mathbf{x} \right|)$.
The isotropic character of the emission allows us to introduce the angle-integrated \emph{ultimate emission rates}
$r_{\nu,{\rm ult}}$ as:
\be
  l_{\nu}(\mathbf{\hat{n}}) =
  \left\{ 
  \begin{array}{r l}
   r_{\nu,{\rm ult}}/\left( 2 \pi \right) & {\rm if} \; \mathbf{\hat{n}} \cdot \mathbf{\hat{n}}_{\tau} \geq 0 \\
   0                                       & \mbox{otherwise}.
  \end{array}
  \right.
  \label{eqn: ultimate rates}
\ee
$r_{\nu,{\rm ult}}$ and $r_{\nu}$ can differ locally, but 
they have to provide the same cooling (spectral) luminosities:
\be
 \int_V \rho \, r_{\nu} \, {\rm d}V  = \int_{V} \rho \, r_{\nu,{\rm ult}} \, {\rm d}V .
 \label{eqn: r_rad integral 1}
\ee
Since $r_{\nu,{\rm ult}}$ represents the ultimate emission rate, {\it after} the diffusion process has drained 
neutrinos from the opaque region to the neutrino surface, $r_{\nu,{\rm ult}} = 0$ inside $V_{\rm thick}$. 
On the other hand, inside $V_{\rm thin}$ diffusion does not take place and $r_{\nu,{\rm ult}} = r_{\nu}$.
In light of this, \refeq{eqn: r_rad integral 1} becomes 
\be
 \int_{V_{\rm thick} \, \cup \, V_{\rm surf}} \rho \, r_{\nu} \, {\rm d}^3 \spc = 
 \int_{V_{\rm surf}} \rho \, r_{\nu,{\rm ult}} \, {\rm d}^3 \spc.
 \label{eqn: r_rad integral 2}
\ee
\refeq{eqn: r_rad integral 2} has a clear physical interpretation: inside $V_{\rm surf}$,  
$r_{\nu,{\rm ult}}$ is obtained 1) from the emission rate, $r_{\nu}$, at the neutrino surface and 
2) from the re-mapping of the emission rates obtained in the opaque region onto the neutrino surface, 
as a consequence of the diffusion process.
A careful answer to this re-mapping problem would rely on the solution of the diffusion equation 
in the optically thick regime and of the Boltzmann equation in the semi-transparent region. 
The ASL algorithm calculates the amount of neutrinos diffusing from a certain volume element.
But it does not provide information about the angular dependence of their flux, neither
about the point of the neutrino surface where they are ultimately emitted. Thus, a phenomenological model is required.
When the properties of the system under investigation 
change on a time-scale larger than (or comparable to)
the relevant diffusion time-scale (see \refsec{sec: preliminary analysis and estimates}), 
the neutrino fluxes can be considered as quasi-stationary.
Under these conditions, the statistical interpretation of the optical depth, as the average number 
of interactions experienced by a neutrino before escaping, suggests to consider 
$\mathbf{\hat{n}}_{\tau}$ 
as the local preferential direction for neutrino fluxes.
While in the (semi-)transparent regime, this unitary vector provides already the
favourite emission direction (see \refeq{eqn: ultimate rates}), 
in the diffusion regime we have to take into account
the spatial variation of $\mathbf{\hat{n}}_{\tau}$.
To this end, at each point $\spc$ in $V_{\rm thick}$, 
we associate a point $\spc_{\rm surf}(\spc)$ in $V_{\rm surf}$
and a related preferential direction
\be
 \mathbf{\hat{n}}_{\rm path}\left( \spc \right) = 
 \frac{ \spc_{\rm surf}(\mathbf{x}) - \spc }{\left| \spc_{\rm surf}(\spc)-\spc \right|},
\ee
according to the following prescription:
the points $\spc$ and $\spc_{\rm surf}$ are connected by a non-straight path $\gamma$ 
that has $\mathbf{\hat{n}}_{\tau}$ as local gradient:
$\gamma(s):\left[0, 1 \right] \rightarrow \left[\spc, \spc_{\rm surf} \right]$, $\spc \in V_{\rm thick}$,
$\spc_{\rm surf} \in V_{\rm surf}$, and ${\rm d} \gamma / {\rm d} s = \mathbf{n}_{\tau}$.
This procedure is sketched in \reffig{fig: path}.\\ 
Once $\mathbf{\hat{n}}_{\rm path}$ has been calculated everywhere inside $V_{\rm thick}$, 
we can re-distribute the neutrinos coming from the optically
thick region on the neutrino surface.
This is done assuming that neutrinos coming from a point $\mathbf{x}$
are emitted preferentially from points of the neutrino surface located around $\spc_{\rm surf}(\spc)$.
More specifically, from points $\spc'$ for which 1) $\spc' \in V_{\rm surf}$; and
2) $\mu(\spc,\spc') \equiv \mathbf{\hat{n}}(\spc,\spc') \cdot \mathbf{\hat{n}}_{\rm path}(\spc) > 0 $, where 
$\mathbf{\hat{n}}(\spc,\spc') \equiv \left( \spc' - \spc \right) / \left| \spc' - \spc \right|$. 
If $\mathbf{\hat{n}}$ and $\mathbf{\hat{n}}_{\rm path}$ are close to the parallel condition (i.e. $\mu \approx 1$) we
expect more neutrinos than in the case of perpendicular directions (i.e. $\mu \approx 0$ ).
We smoothly model this effect assuming a $\mu^2$ dependence. \\
The global character of this re-mapping procedure
represents a severe computational limitation for our large, three dimensional, MPI-parallelised Cartesian
simulation. In order to make the calculation feasible, 
we take again advantage of the expected high degree of axial symmetry of remnant 
(especially in the innermost part
of it, where the diffusion takes place and most of the neutrino are emitted), 
and we compute $r_{\nu,{\rm ult}}$ in axisymmetry.

%
%
\subsection{Initial Conditions}
\label{sec: initial conditions}

\begin{figure*}
\begin{center}
\includegraphics[width = \linewidth]{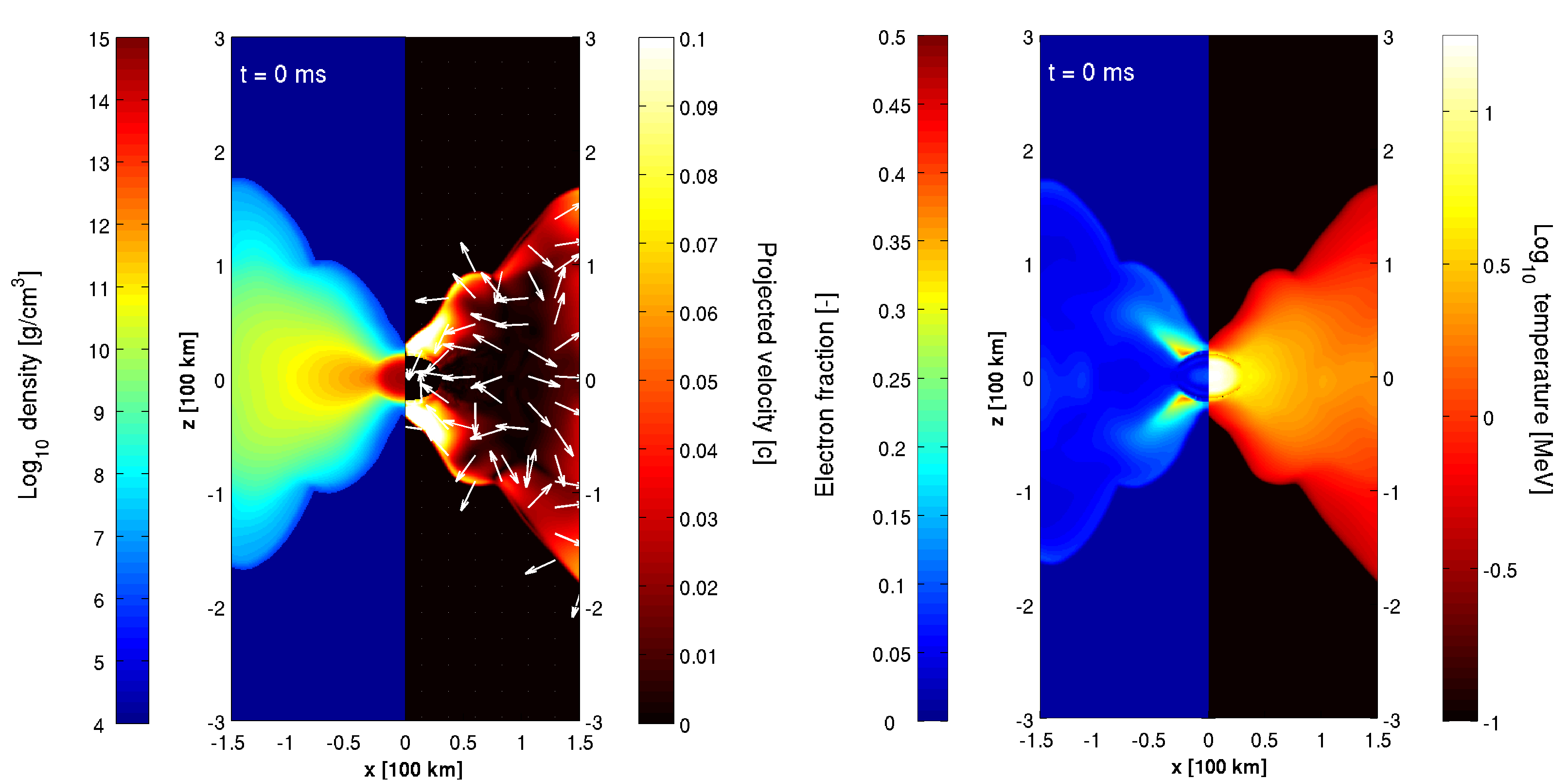}
\end{center}
\caption{Vertical slices of the three dimensional domain (corresponding to the $y=0$ plane), recorded at
the beginning of the simulation. In the left panel, we color coded the logarithm of the matter density (in ${\rm g/cm^3}$, left side) 
and the projected fluid velocity (in units of $c$, on the right side); 
the arrows indicate the direction of the projected velocity in the plane). On the right panel, we represent the
electron fraction (left side) and the logarithm of the matter temperature (in unit of MeV, right side).}
\label{fig: initial conditions}
\end{figure*}

The current study is based on previous, 3D hydrodynamic studies of the merger of two
non-spinning 1.4 \msun neutron stars. This simulation was performed with a 
3D Smoothed Particle Hydrodynamics (SPH) code, the implementation details of which 
can be found in the literature \citep{Rosswog2000,Rosswog2003,Rosswog2005,
Rosswog2007c}. For overviews over the SPH method, the interested reader is referred to recent reviews 
\citep{Monaghan2005,Rosswog2009b,Springel2010a,Price2012a,Rosswog2014b,Rosswog2014c}.
The neutron star matter is modelled with the Shen et al. EoS \citep{Shen1998a,Shen1998b}, 
and the profiles of the density and $\beta$-equilibrium electron fraction can be found in fig. 1 
of \cite{Rosswog2013}.
During the merger process the debris can cool via neutrino emission, and electron/positron captures
can change the electron fraction. These processes are included via the
opacity-dependent, multi-flavor leakage scheme of \cite{Rosswog2003}. Note, however, 
that no heating via neutrino absorption is included. Their effects are
the main topic of the present study.\\
As the starting point of our neutrino-radiation hydrodynamics study, we consider the matter distribution
of the 3D SPH simulation with $10^6$ particles, at 15 ms after the first contact  
(corresponding to 18 ms after the simulation start).
Not accounting for the neutrino absorption during this short time, should only have a 
small effect, since, according to the estimates from \refsec{sec: preliminary analysis and estimates},
the remnant hardly had time to change.\\
We map the 3D SPH matter distributions of density, temperature, electron fraction 
and fluid velocity on the Cartesian, equally spaced
grid of \texttt{FISH}, with a resolution of 1 {\rm km}.
The initial extension of the grid is 
$\left( 800 {\rm km} \times 800 {\rm km} \times 640 {\rm km} \right)$.
During the simulation, we increase the domain in all directions to follow the wind expansion,
keeping the HMNS always in the centre.
At the end, the computational box is 
$\left( 2240 {\rm km} \times 2240 {\rm km} \times 3360 {\rm km} \right)$ wide.\\
The initial data cover a density range of 
$ 10^8 {\rm g \, cm^{-3}} - 3.5 \times 10^{14} {\rm g \, cm^{-3}}$.
Surrounding the remnant, we place an inert atmosphere, characterised
by the following stationary properties:
$\rho_{\rm atm} = 5 \cdot 10^{3}\, {\rm g/cm^3}$, $T_{\rm atm} = 0.1
\, {\rm MeV}$, $Y_{e,{\rm atm}}=0.01$ and $\mathbf{v}_{\rm atm} =
\mathbf{0}$.
The neutrino source terms are set to 0 in this atmosphere. With this treatment, we
minimize the influence of the atmosphere on the disc and on the wind dynamics. 

Even though in our model we try to stay as close as possible to the choices adopted in the SPH simulation,
initial transients appear at the start of the simulation.
One of the causes is the difference in the spatial resolutions between the two models.
The resolution we are adopting in \texttt{FISH} is significantly lower than the one 
provided by the initial SPH model inside the HMNS, $\sim 0.125 \, {\rm km}$, (which is
necessary to model consistently the central object), while it is comparable or better inside the disc.
Due to this lack of resolution, we decide 
to treat the HMNS as a stationary rotating object.
To implement this, we perform axisymmetric averages 
of all the hydrodynamical quantities 
at the beginning of the simulation. 
At the end of each hydrodynamical time step, we re-map these profiles
in cells contained inside an ellipsoid, 
with $a_x = a_y = 30 \, {\rm km}$ and $a_z = 23 \, {\rm km}$, 
and for which $\rho > 2 \cdot 10^{11} {\rm g/cm^3}$.
For the velocity vector, we consider only the azimuthal component,
since 1) the HMNS is rotating fast around its polar axis (with a period $P \approx 1.4 \, {\rm ms}$)
and 2) the non-azimuthal motion inside it is characterised by much smaller velocities (for example,
$\left| v_R \right| \sim 10^{-3} \left| v_\phi \right| $, where $v_R$ and $ v_\phi $ are the radial and the
azimuthal velocity components).
Concerning the density and the rotational velocity profiles, our treatment is consistent
with the results obtained by \cite{Dessart2009} (fig. 4), who showed that
$\sim 100 \, {\rm ms}$ after the neutron star have collided
those quantities have changed only slightly inside the HMNS.
We expect the electron fraction and the temperature also to stay close to their initial values,
since the most relevant neutrino surfaces for $\nue$ and $\nueb$ 
are placed outside the stationary region and the diffusion time-scale is much longer than the simulated
time (see, for example, \refsec{sec: preliminary analysis and estimates}).

\begin{figure}
\includegraphics[width=\linewidth]{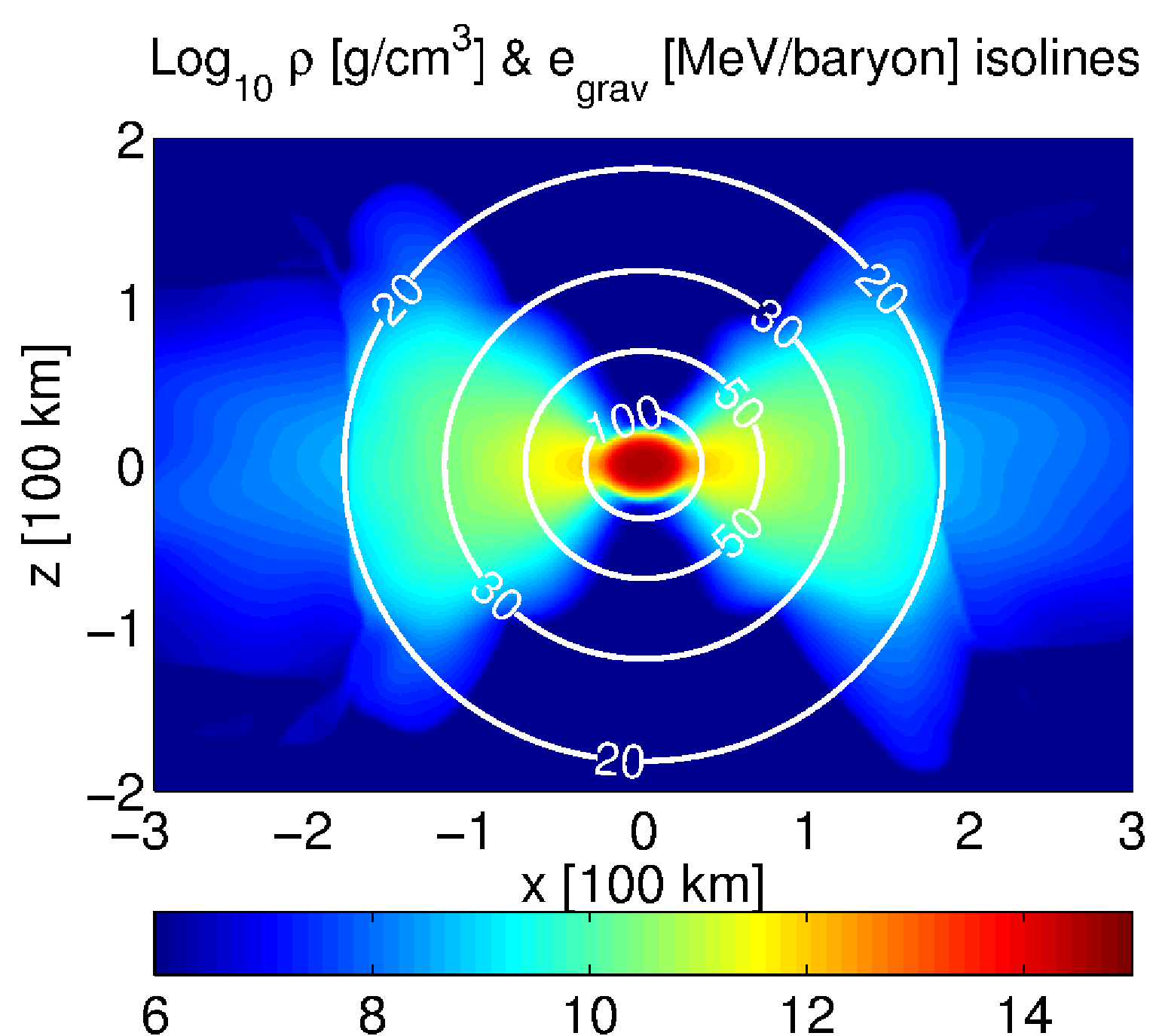}
\caption{Logarithm of the matter density (color coded, in ${\rm g \, cm^{-3}}$) 
and isocontours of the gravitational energy (white lines, in ${\rm MeV \, baryon^{-1}}$),
on a vertical slice of the three dimensional domain, at $t=0$.}
\label{fig: rho and grav}
\end{figure}

To give the opportunity to the system to adjust to a more stable configuration on the new grid,
we consider the first $ 10 \, {\rm ms}$ of the simulation as a ``relaxation phase''.
During this phase, we evolve the system considering only neutrino emission.
Its duration is chosen so that the initial transients arrive at 
the disc edge, and the profiles inside the disc reach new quasi-stationary 
conditions. The ``relaxed'' conditions are visible in \reffig{fig: initial conditions}. 
They are considered as the new initial conditions and
we evolved them for $\sim 90 \, {\rm ms}$, including the 
effect of neutrino absorption.
In the following, the time $t$ will be measured with respect to this second re-start.
During the relaxation phase, we notice an increase of the electron 
fraction, from 0.05 up to 0.1-0.35, for a tiny amount of matter ($\la 10^{-5}$ \msun) in the low density region 
($\rho \la 10^9 {\rm g/cm^3}$) situated above the 
innermost, densest part of the disc ($R_{\rm cyl} \la 50 \, {\rm km}$, $\left| z \right| \ga 20 \, {\rm km}$). 
Here, the presence of neutron-rich, hot matter
in optically thin conditions favours the emission of $\nueb$, via
positron absorption on neutrons.
A similar increase of $Y_e$ is also visible in the original SPH simulations,
for times longer than 15 ms after the first collision.\\
In \reffig{fig: rho and grav} we show isocontours of the absolute value of gravitational specific energy,
drawn against the colour-coded matter density, at the beginning of our simulation.
The gravitational energy provides an estimate of the energy that neutrinos have to deposit to unbound matter,
at different locations inside the disc (see \refsec{sec: preliminary analysis and estimates}).

%
%

\section{Simulation results}
\label{sec: results}

%
%
\subsection{Disc evolution and matter accretion}
\label{sec: disc evolution}

\begin{figure}
\begin{center}
\includegraphics[width = \linewidth]{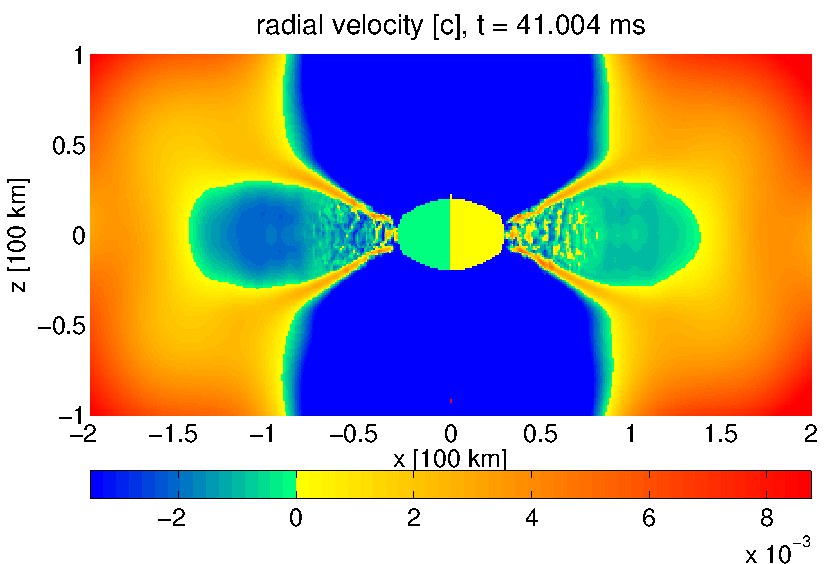}
\end{center}
\caption{Vertical slice of the inner part of the three dimensional domain ($y=0$ plane), 
taken at $41 \, {\rm ms}$ after the beginning 
of the simulations. Color coded is the radial component of the fluid velocity.
The two coloured hemispheres in the centre represent the stationary central object for which $v_r \approx 0$ 
(the two actual colours are very small numbers).}
\label{fig: vrad_acc}
\end{figure}

After the highly dynamical merger phase, the remnant is still dynamically evolving and
not yet in a perfectly stationary state.\\
In \reffig{fig: vrad_acc}, we show the radial component 
of the fluid velocity on the $y=0$ plane, at $ 41 \, {\rm ms}$ after the beginning of the simulation.
The central part of the disc, corresponding to a density contour of
$\sim 5 \cdot 10^9 \, {\rm g \, cm^{-3}}$, is slowly being
accreted onto the HMNS ($v_R \sim$ a few $10^{-3}c$), while the outer edge is gradually expanding 
along the equatorial direction. The velocity profile shows interesting asymmetries and deviations 
from an axisymmetric behaviour.
The surface of the HMNS and the innermost part of the disc are characterised by steep gradients of
density and temperature, and they behave like a pressure wall for the infalling matter. Outgoing
sound waves are then produced and move outwards inside the disc, transporting energy, linear
and angular momentum.
At a cylindrical radius of $R_{\rm cyl} \la 80 \, {\rm km}$, they induce 
small scale perturbations in the velocity field, visible as bubbles of slightly positive
radial velocity. 
These perturbations dissolve at larger radii, releasing their momentum and energy inside the disc, and
favouring its equatorial expansion.

\begin{figure}
\begin{center}
\includegraphics[width = \linewidth]{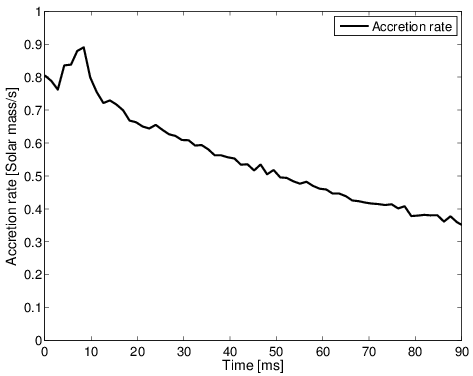}
\end{center}
\caption{Temporal evolution of the accretion rate on the HMNS, $\dot{M}$, calculated
as the net flux of matter crossing a cylindrical surface of radius 
$R_{\rm cyl}=35 \, {\rm km}$ and axis corresponding to the rotational axis of the disc.}
\label{fig: accretion_eta_time}
\end{figure} 

\noindent The temporal evolution of the accretion rate $\dot{M}$,
computed as the net flux of matter crossing a cylindrical surface of radius 
$R_{\rm cyl}=35 \, {\rm km}$ and axis corresponding to the rotational axis of the disc,
is plotted in \reffig{fig: accretion_eta_time}.
This accretion rate is compatible with the estimate performed in 
\refsec{sec: preliminary analysis and estimates} using an $\alpha$-viscosity disc model. 
A direct comparison with \refeq{eqn: mdot} suggests
an effective parameter $\alpha \approx 0.05$ for our disc. 
We stress again that no physical viscosity
is included in our model: the accretion is driven by unbalanced pressure gradients, 
neutrino cooling (see \refsec{sec: neutrino emission}) 
and dissipation of numerical origin.
However, the previous estimate is useful to compare our disc with
purely Keplerian discs, in which a physical $\alpha$-viscosity has been included 
(usually, with $0.01 \la \alpha \la 0.1$). Our value of $\alpha \approx 0.05$ is close to
what is usually assumed for such discs ($\sim 0.1$). Higher viscosities would enhance
the neutrino emission and probably the mass loss.\\

\begin{figure*}
\begin{center}
\includegraphics[width = \linewidth]{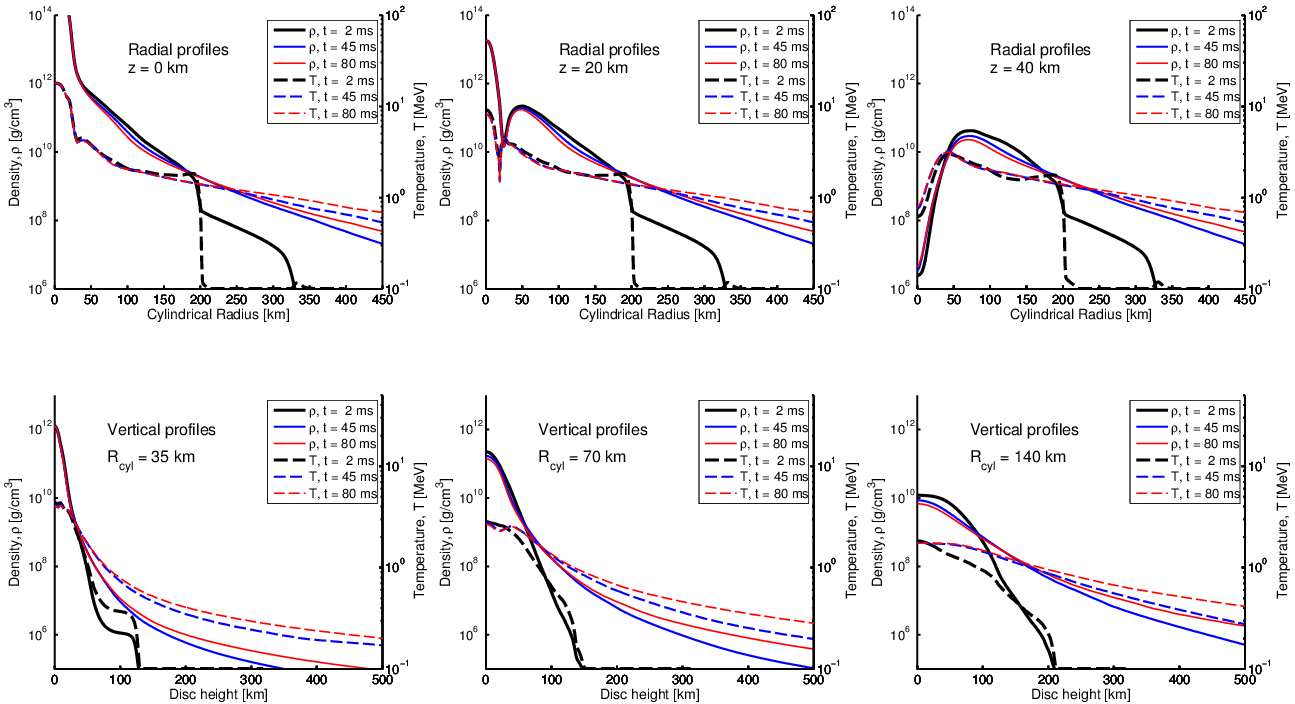}
\end{center}
\caption{Radial (upper row) and vertical (lower row) profiles of the axisymmetric density (solid lines) 
and temperature (dashed lines)
inside the disc, recorded at different times during the simulation ($t \approx 2 \, {\rm ms}$ (black-thick lines), 
$t \approx 45 \, {\rm ms}$ (blue-normal lines),
$t \approx 80 \, {\rm ms}$ (red-thin lines)). 
The different columns correspond to different values of the section coordinates: from left to right,
$ z = 0 \, {\rm km}, \, 20 \, {\rm km}, \, 40 \, {\rm km}$ for the radial profiles; 
$R_{\rm cyl} = 35 \, {\rm km}, \, 70 \, {\rm km}, \, 140 \, {\rm km}$ for the vertical ones.}
\label{fig: disc profiles}
\end{figure*}

On a timescale of a few tens of milliseconds, the profiles inside the disc change, as
consequence of the accretion process and of the outer edge expansion.
These effects are visible in the upper row of \reffig{fig: disc profiles}, where
radial profiles of temperature and density are drawn, at different times and heights
inside the disc.
We notice, in particular, that the density decreases in the internal part of the disc 
($50 \, {\rm km} \la R_{\rm cyl} \la 200 \, {\rm km}$),
as result of the accretion.
In the same region, the balance between 
the increase of internal energy and the efficient cooling provided by neutrino emission
keeps the temperature almost stationary.
At larger radial distances ($R_{\rm cyl} \ga 200 \, {\rm km}$), 
the initial accretion of a cold, thin layer of matter (visible in the $t=2 \, {\rm ms}$ profiles) is followed
by the continuous expansion of the outer margin of the hot internal disc.

%
%
\subsection{Neutrino emission}
\label{sec: neutrino emission}


\begin{figure*}
\begin{center}
\includegraphics[width = \linewidth]{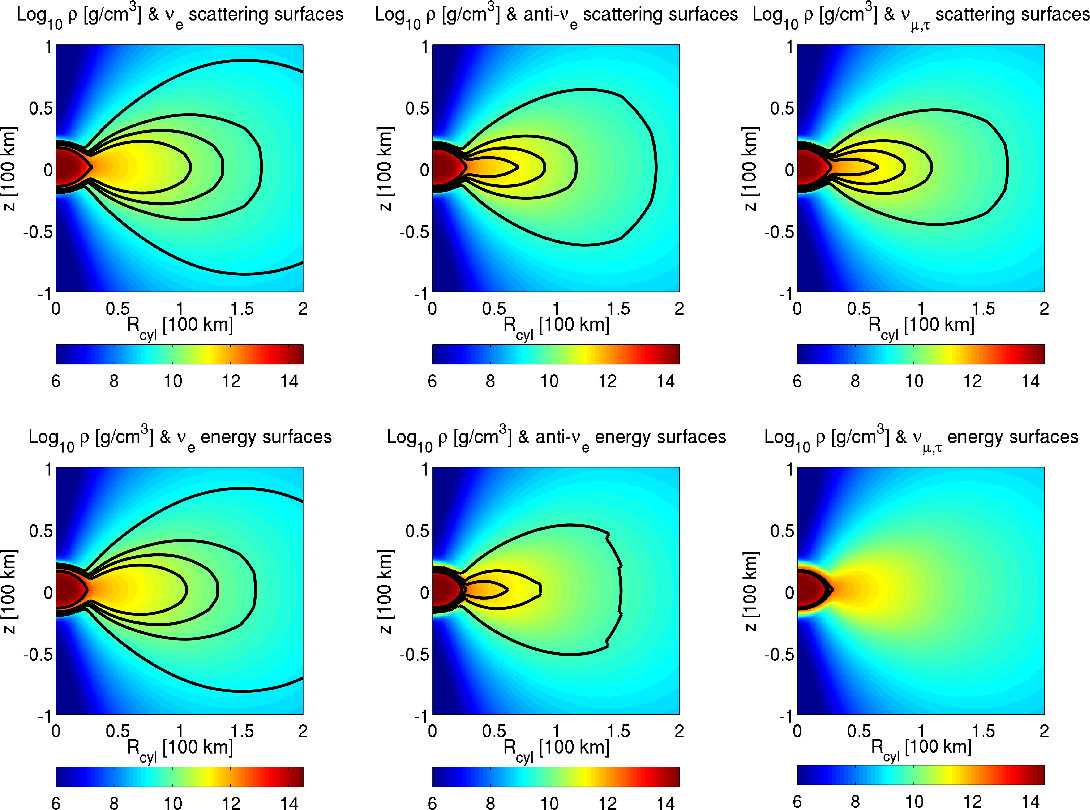}
\end{center}
\caption{Location of the neutrino surfaces for $\nue$ (left column), $\nueb$ (central column) and $\numt$ (right column),
for the scattering optical depth (upper row) and for the energy optical depth (bottom row), $40 \, {\rm ms}$ after
the beginning of the simulation.
Color coded is the logarithm of cylindrically averaged matter density, $\rho \, [{\rm g/cm^3}]$. 
The different lines correspond to the neutrino surfaces for
different values of the neutrino energy: from the innermost line to the outermost one, $E_{\nu} = 
4.62 \, {\rm MeV}, \, 10.63 {\rm MeV}, \, 16.22 {\rm MeV}, \, 24.65 {\rm MeV}, \, 56.96 {\rm MeV} $. }
\label{fig: nu_surf_tot}
\end{figure*}

\begin{figure}
 \begin{center}
  \includegraphics[width = \linewidth]{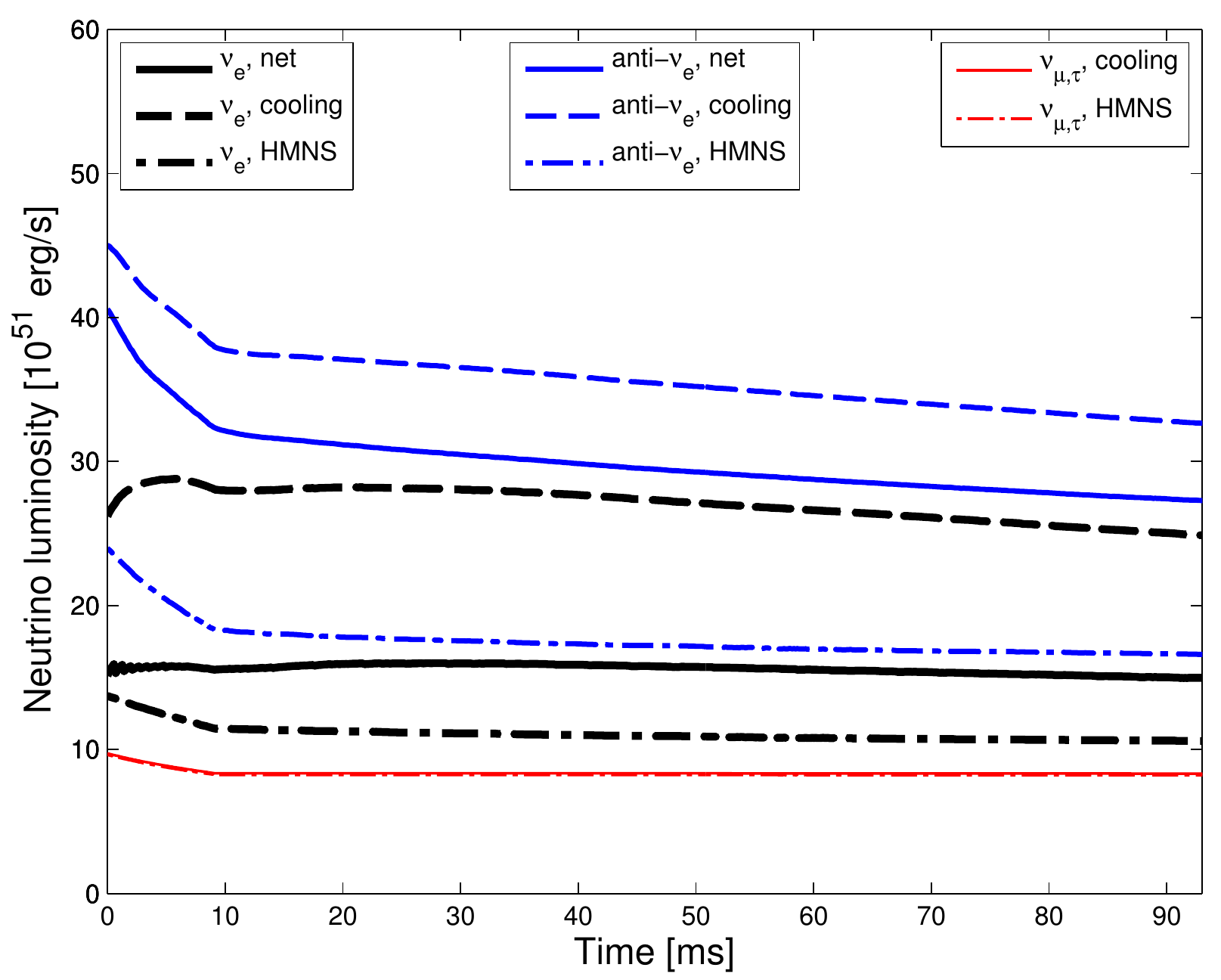}
 \end{center}
 \caption{Time evolution of the net (solid) and cooling (dashed) luminosities obtained 
by the ASL scheme for $\nue$ (black-thick), $\nueb$ (blue-normal)
and $\numt$ (red-thin) neutrino species. The difference between the cooling and the net luminosities is
represented by the re-absorbed luminosity. The contributions to the cooling luminosities
coming from the HMNS, defined as the volume characterised by $\rho > 5 \cdot 10^{11} {\rm g \, cm^{-3}}$,
is also plotted (dot-dashed lines).
Note that for $\numt$, the net and the cooling luminosities coincide, 
and they are almost equal to the HMNS contribution.}
\label{fig: lum with time}
\end{figure}

\begin{figure}
\begin{center}
\includegraphics[width = \linewidth]{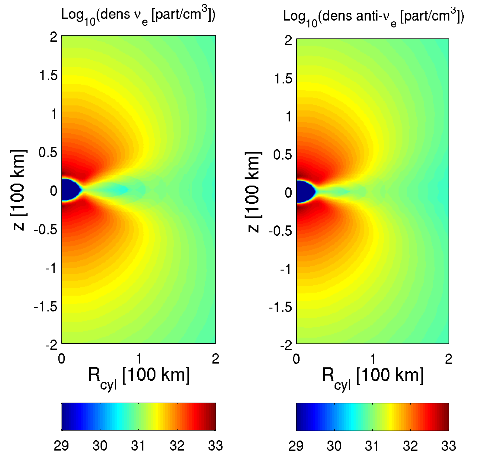}
\end{center}
\caption{Energy-integrated (axisymmetric) neutrino density of the free-streaming neutrinos, 
$N_{\nu}$, for $\nue$ (left panel) and $\nueb$ (right panel), calculated outside the innermost neutrino surface 
(corresponding to $E_{\nu} = 3 \, {\rm MeV}$), at $t \approx 40 \, {\rm ms}$ after the beginning of the simulation.}
\label{fig: nu dens}
\end{figure}

\begin{figure}
\begin{center}
\includegraphics[width = \linewidth]{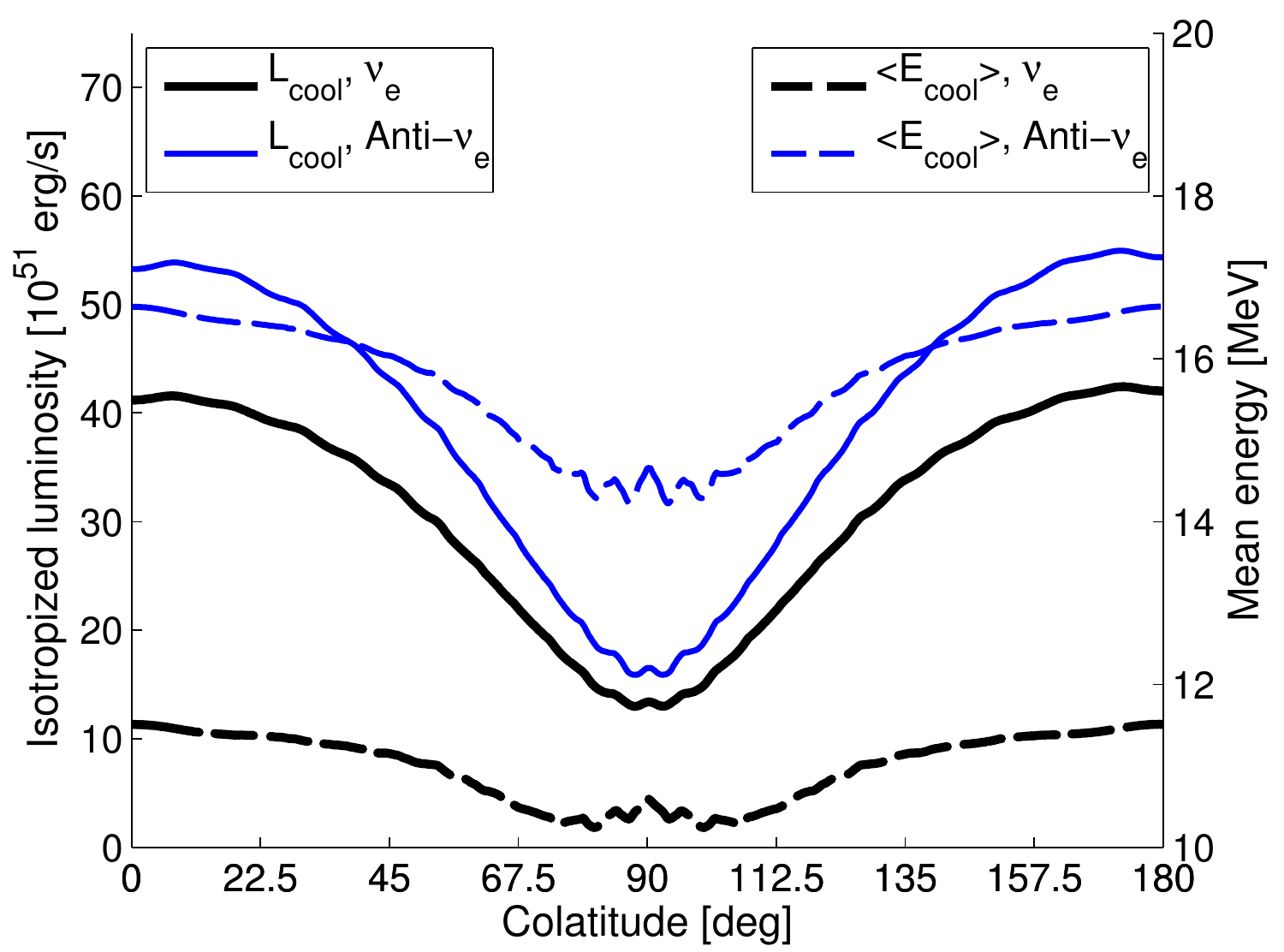}
\end{center}
\caption{Angular dependence of the isotropised neutrino cooling luminosities (solid line) 
and of the neutrino mean energies (dashed lines), as a function of the colatitude.
The black-thick lines correspond to $\nue$, while the blue-thin lines to $\nueb$. 
As a representative time, we consider $t \approx 40 {\rm ms}$ after the 
beginning of the simulation.}
\label{fig: ang lum aven}
\end{figure}

In \reffig{fig: nu_surf_tot}, we show the neutrino
surfaces obtained by the calculation of the spectral neutrino optical depths, 
together with the matter density distribution (axisymmetric, color coded).
Different lines correspond to different energy bins.
In the upper panels, we represent the scattering neutrino surfaces, while in the lower panels
the energy ones. Their shapes follow closely the matter density distribution,
due to the explicit dependence appearing in \refeq{eqn: tau cl def} and \refeq{eqn: tau eff def}. 
The last scattering surfaces for the 
energies that are expected to be more relevant for the neutrino emission 
($10 \, {\rm MeV} \la E_{\nu} \la 25 \, {\rm MeV} $, corresponding to the expected range for the mean energies,
as we will discuss below) extend far outside in the disc, 
compared with the radius of the central object.
$\nue$'s have the largest opacities, due to the extremely neutron rich environment that 
favours processes like neutrino absorption on neutrons.
Since the former reaction is also very efficient in thermalising neutrinos, 
the scattering and the energy neutrino
surfaces are almost identical for $\nue$'s.
In the case of $\nueb$'s, the relatively low density of free protons
determines the reduction of the scattering and, even more, of the energy optical depth.
For $\numt$'s, neutrino bremsstrahlung and $e^+-e^-$ annihilation freeze out
at relatively high densities  
and temperatures ($\rho \sim 10^{13} \, {\rm g/cm^3}$ and $k_{\rm B}T \sim 8 \, {\rm MeV}$),
reducing further the energy neutrino surfaces, while elastic scattering on nucleons
still provides a scattering opacity comparable to the one of $\nueb$'s.

The energy- and volume-integrated luminosities obtained during the simulation are presented 
in \reffig{fig: lum with time}.
The cooling luminosities for $\nue$'s and $\nueb$'s (dashed lines)
decrease weakly and almost linearly with time.
This behaviour reflects the continuous supply of hot accreting matter.
The faster decrease of $\dot{M}$ (cf. \reffig{fig: accretion_eta_time}) 
would imply a similar decrease in the luminosities, if the neutrino radiative efficiency
of the disc were constant. However, the latter increases with time due to the decrease of density and
the constancy of temperature characterising the innermost part of the disc 
(see \refsec{sec: disc evolution}).
Also the luminosity for the $\numt$ species is almost constant. 
This is a consequence of the stationarity
of the central object, since most of the $\numt$'s come from there. However, this result
is compatible with the long cooling time-scale of the HMNS, \refeq{eqn: diffusion time HMNS}.
We specify here that the plotted lines for $\numt$ correspond to one single species. Thus, the
\emph{total} luminosity coming from heavy flavour neutrinos is four times the plotted one, see also
\refeq{eqn: edot contributions}.\\
In the case of $\nue$'s and $\nueb$'s, the luminosity provided by $V_{\rm HMNS}$
(defined in \refsec{sec:neutrino treatment} and represented by dot-dashed lines in \reffig{fig: lum with time})
and the luminosity of the accreting disc are comparable. This result is compatible with what is observed 
in core collapse supernova simulations \citep[see, for example, fig. 6 of][]{Liebendorfer2005a}, a few tens of milliseconds after bounce:
assuming a density cut of $5 \times 10^{11} {\rm g \, cm^{-3}}$ 
for the proto-neutron star, its contribution
is roughly half of the total emitted luminosity, for both $\nue$ and $\nueb$.
Instead, if we further restrict $V_{\rm HMNS}$ only to the central ellipsoid
(see \refsec{sec: initial conditions} for more details),
we notice that the related luminosity reduces to $\la 10 \times 10^{51} {\rm erg \, s^{-1}}$
for all neutrino species. This is consistent with our preliminary estimate, 
\refeq{eqn: HMNS luminosity}.\\
The inclusion of neutrino absorption processes in the optically thin region 
reduces the cooling luminosities to the net luminosities 
(solid lines in \reffig{fig: lum with time}).
For $\nue$'s, the neutron rich environment reduces the number luminosity by $\approx 37$ per cent, while
for $\nueb$'s this fraction drops to $\approx 14$ per cent.

The values of the neutrino mean energies are practically stationary during the simulation:
from the net luminosities at $t \approx 40 {\rm ms}$, 
$\langle E_{\nue} \rangle \approx 10.6 \, {\rm MeV}$, 
$\langle E_{\nueb} \rangle \approx 15.3 \, {\rm MeV}$
and $\langle E_{\numt} \rangle \approx 17.3 \, {\rm MeV}$. 
The mean neutrino energies show the expected hierarchy, 
$\langle E_{\nue} \rangle < \langle E_{\nueb} \rangle < \langle E_{\numt} \rangle$, 
reflecting the different locations of the thermal decoupling surfaces.
While the values obtained for $\nue$'s and $\nueb$'s are consistent with previous calculations, 
$\langle E_{\numt} \rangle$ is smaller than expected \citep[see, for example,][]{Rosswog2013}. 
This is due to the lack of resolution at the HMNS surface, 
where most of the energy neutrino surfaces for $\numt$ are located.
This discrepancy has no dynamical effects for us, since most of $\numt$ come from the 
stationary central object.

The ray-tracing algorithm, see \refsec{sec:neutrino treatment}, allows us to compute 
1) the neutrino densities outside the neutrino surfaces; 
2) the angular distribution of the isotropised neutrino cooling luminosities and
mean neutrino energies, as seen by a far observer.
In \reffig{fig: nu dens}, we represent the energy-integrated 
axisymmetric neutrino densities $N_{\nu}$, \refeq{eqn: nu_dens_eq}, 
for $\nue$ (left) and $\nueb$ (right). 
These densities reach their maximum in the funnel above the HMNS,
due to the geometry of the emission and to the short distance from the most emitting regions.
At distances much 
larger than the dimension of the neutrino surfaces, $N_{\nu}$ shows the expected $R^{-2}$ dependence.\\
The disc geometry introduces a clear anisotropy in the neutrino emission,
visible in \reffig{fig: ang lum aven}.
Due to the larger opacity along the equatorial direction, the isotropic luminosity along
the poles is $\sim 3 - 3.5$ more intense than the one along the equator.
The different temperatures at which neutrinos decouple from matter at different polar angles
determine the angular dependence of the mean energies.

%
%
\subsection{Neutrino-driven wind}

\begin{figure*}
\begin{center}
\includegraphics[width = \linewidth]{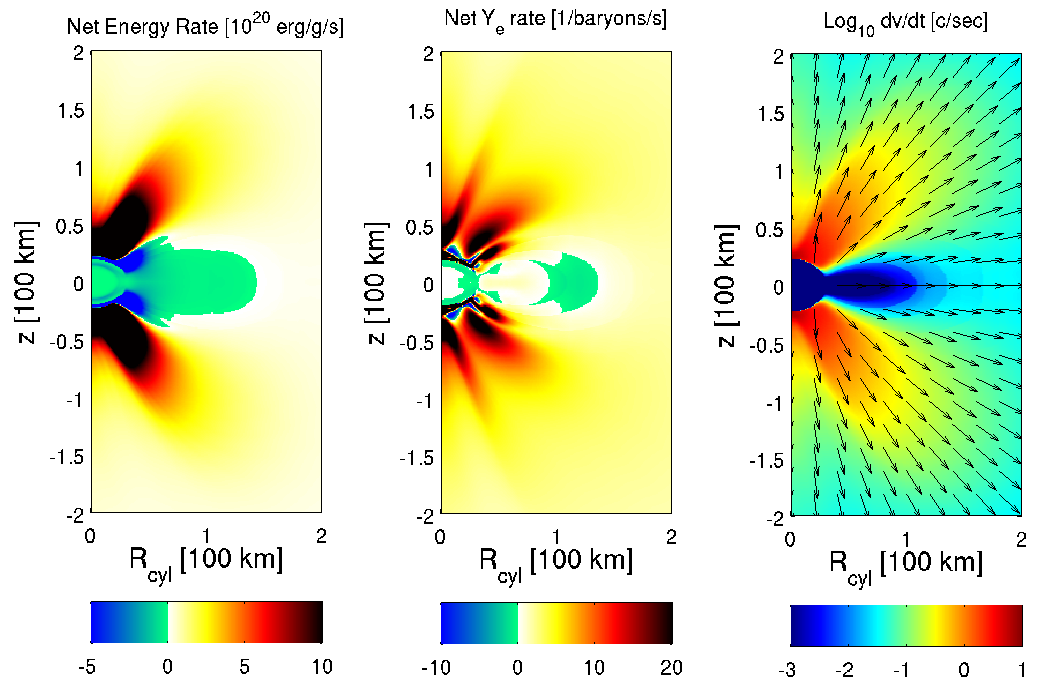}
\end{center}
\caption{Energy- and species-integrated axisymmetric $\nu$ net rates for energy 
(left panel, in units of $10^{20}$ erg/g/s) and $Y_e$ (central panel, in units of 1/baryon/s), 
and of the fluid velocity variation provided by neutrino absorption in the optically thin regions 
(right panel, in units of $c/{\rm s}$). 
As a representative time, we consider $t \approx 40 \, {\rm ms}$ after the beginning of the simulation.
The complex structure of the net $Y_e$ rate in the funnel, above the HMNS poles, 
originates from the variety of conditions of $Y_e$, $\rho$ and $\mathbf{v}$ at that specific moment.}
\label{fig: net rates}
\end{figure*}

The evolution of the disc and the formation of a neutrino-driven wind depend crucially on the 
competition between neutrino emission and absorption. In \reffig{fig: net rates}, 
we show axisymmetric averages of the net specific energy rate (left), 
of the net electron fraction rate (centre), and of
the acceleration due to neutrino absorption (right), at $t=40 \, {\rm ms}$.\\
Inside the most relevant neutrino surfaces and a few kilometers outside them, neutrino cooling
dominates. Above this region, neutrino heating is always dominant. 
The largest neutrino heating rate happens in the funnel,
where the neutrino densities are also larger. However, these regions are characterised
by matter with low density ($\rho \la 10^{7} {\rm g \, cm^{-3}}$) and small specific angular momentum.
Thus, this energy deposition has
a minor dynamical impact on this rapidly accreting matter. 
On the other hand, at larger radii 
($80 \, {\rm km} \la R_{\rm cyl} \la 120 \, {\rm km}$)
net neutrino heating affects denser matter ($\rho \la 10^{10}{\rm g \, cm^{-3}}$),
rotating inside the disc around the HMNS. This combination provides an efficient net energy deposition.\\
Neutrino diffusion from the optically thick region determines small variations around 
the initial weak equilibrium value in the electron fraction.
On the contrary, in optically thin conditions, the initial very low electron fraction 
favours reactions like the absorption of $e^+$ and 
$\nue$ on free neutrons. Both processes
lead to a positive and large $\left({\rm d}Y_e/{\rm d}t \right)_{\nu}$, in association with efficient energy deposition.\\
Due to the geometry of the emission and to the shadow effect
provided by the disc,
the direction of the acceleration provided by neutrino absorption is approximately radial,
but its intensity is much larger in the funnel, where the energy deposition is also more intense.

As a consequence of the continuous neutrino energy and momentum deposition,
the outer layers of the disc start to expand a few milliseconds after the beginning of the simulation, 
and they reach an almost stable configuration in a few tens of milliseconds.
Around $t \sim 10 \, {\rm ms}$, also the neutrino-driven wind starts to develop from the expanding disc.
Wind matter moves initially almost vertically (i.e., with
velocities parallel to the rotational axis of the disc), 
decreasing its density and temperature during the expansion.
We show the corresponding vertical profiles inside the disc
in the bottom panels of \reffig{fig: disc profiles},
at different times and for three cylindrical radii.
Both the disc and the wind expansions are visible in the rise of the density and
temperature profiles, especially at cylindrical radii of $70 \, {\rm km}$ and $140 \, {\rm km}$.\\
Among the energy and the momentum contributions, the former is the most important one
for the formation of the wind.
To prove this, we repeat our simulation in two cases, starting from
the same initial configuration and relaxation procedure. In a first case, we set the heating
rate $h_{\nu}$ appearing in \refeq{eqn: particle heating rate} and \refeq{eqn: stress rate}
to 0. Under this assumption, we observe neither the disc expansion nor the wind formation.
In a second test, we include the effect of neutrino absorption only in the energy and $Y_e$
equations, but not in the momentum equation. In this case, the wind still develops and 
its properties are qualitatively very similar to our reference simulation.\\

\begin{figure*}
\begin{center}
\includegraphics[width = \linewidth]{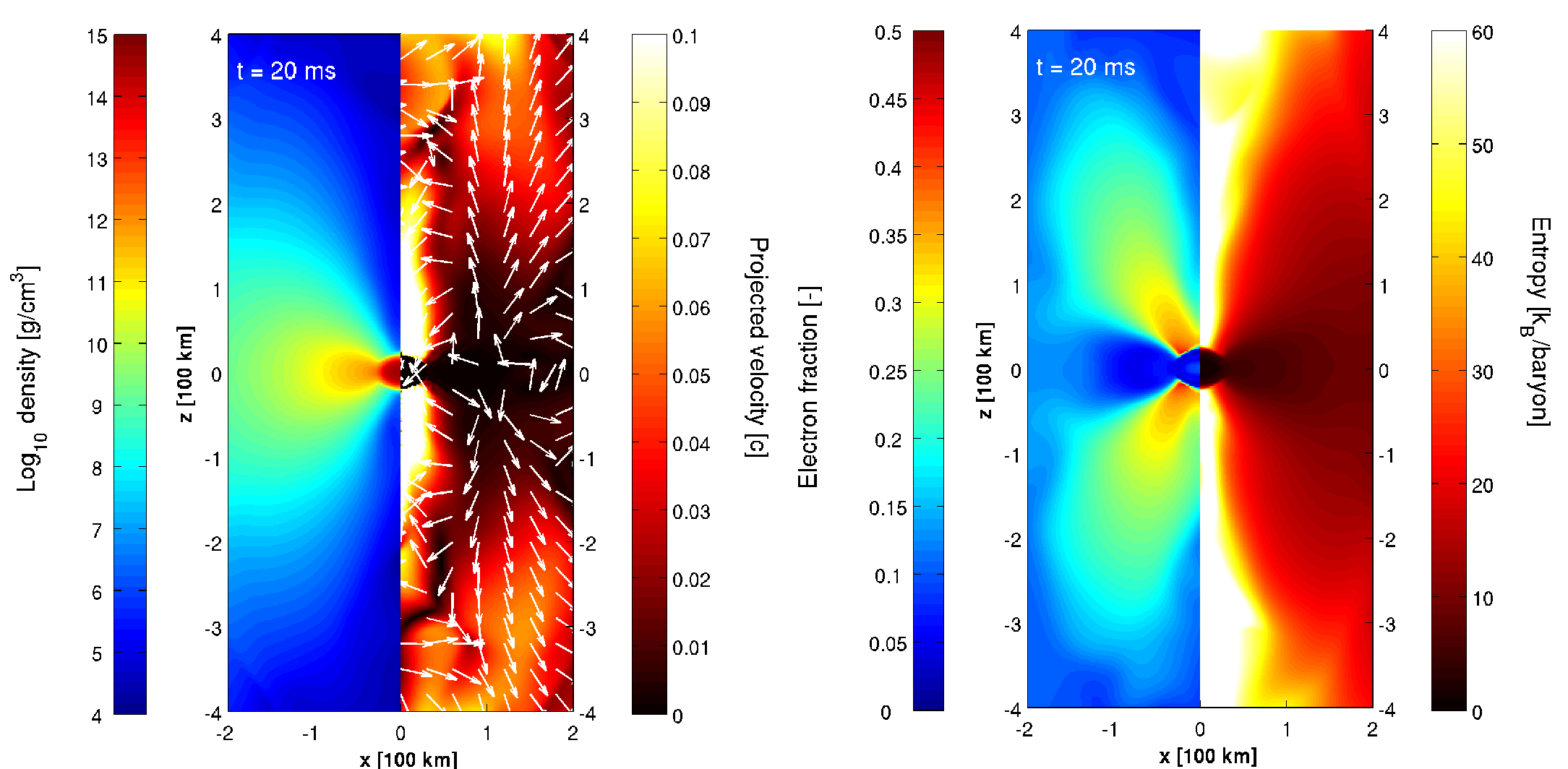}
\end{center}
\caption{Vertical slices of the three dimensional domain (corresponding to the $y=0$ plane), recorded $20 \, {\rm ms}$ after
the beginning of the simulation. 
In the left panel, we represent the logarithm of the matter density (in ${\rm g/cm^3}$, left side) 
and the projected fluid velocity (in units of $c$, on the right side); 
the arrows indicate the direction of the projected velocity in the plane). On the right panel, we represent the
electron fraction (left side) and the matter entropy (in unit of $k_{\rm B}/{\rm baryon}$, right side).}
\label{fig: wind 20ms}
\end{figure*}

\begin{figure*}
\begin{center}
\includegraphics[width = \linewidth]{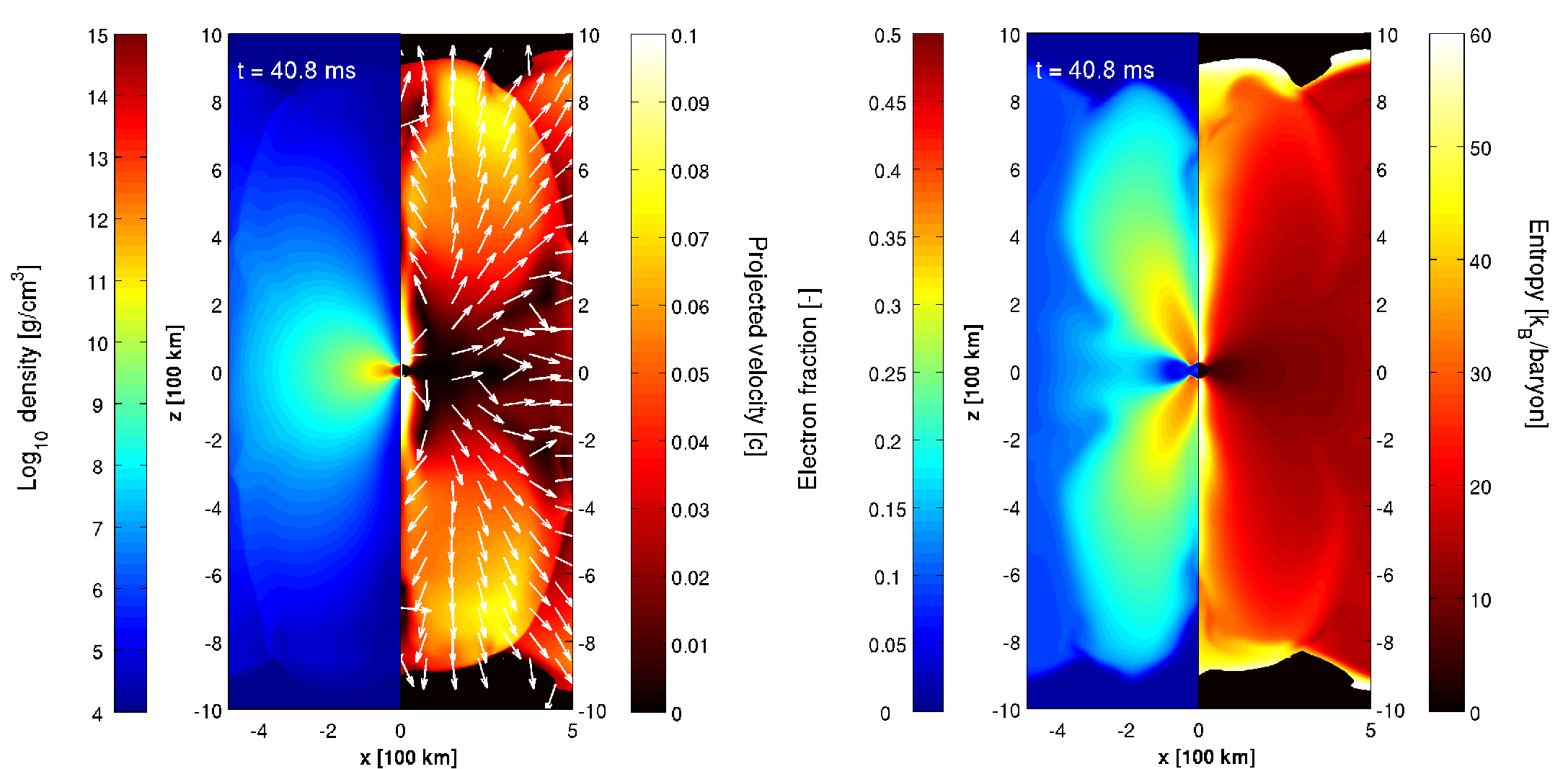}
\end{center}
\caption{Same as in \reffig{fig: wind 20ms}, but at $\approx 40 \, {\rm ms}$ after the beginning of the simulation.}
\label{fig: wind 40ms}
\end{figure*}

\begin{figure*}
\begin{center}
\includegraphics[width = \linewidth]{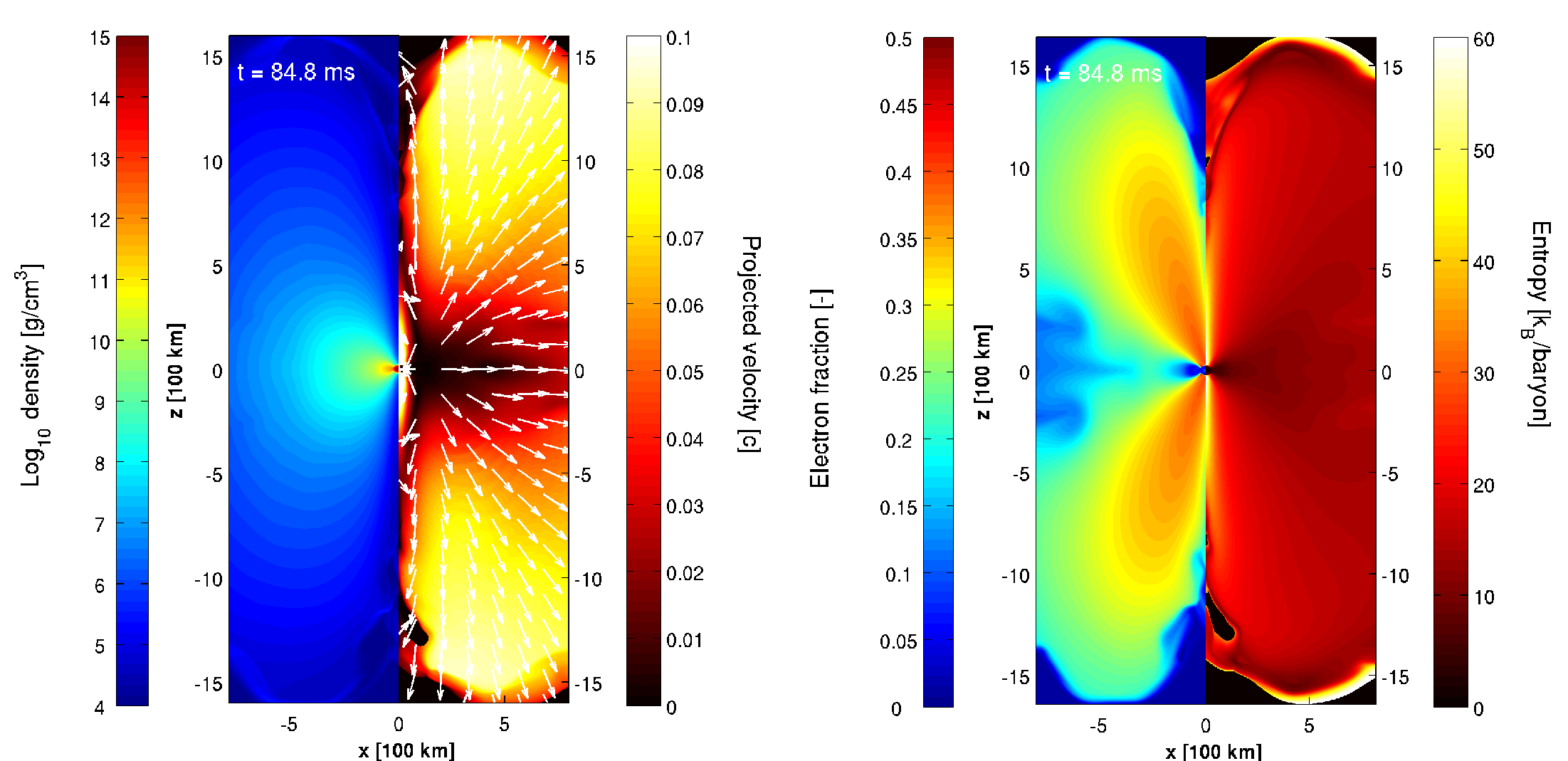}
\end{center}
\caption{Same as in \reffig{fig: wind 20ms}, but at $\approx 85 \, {\rm ms}$ after the beginning of the simulation.}
\label{fig: wind 85ms}
\end{figure*}

In Fig. \ref{fig: wind 20ms}, \ref{fig: wind 40ms} and \ref{fig: wind 85ms} we present 
three different times of the wind expansion,  
$t=20,40,85 \, {\rm ms}$.
To characterise them, we have chosen vertical slices of the three dimensional domain,
for the density and the projected velocity (left picture), and for the electron fraction and the matter entropy (right picture).\\
The development of the wind is clearly associated with the progressive increase of the electron fraction.
The resulting $Y_e$ distribution is not uniform,
due to the competition between the wind
expansion time-scale (\refeq{eqn: wind time-scale}) and the time-scale for weak equilibrium to establish.
The latter can be estimated as $t_{\rm weak} \sim Y_{e,{\rm eq}} / \left( {\rm d}{Y_e}/ {\rm d} t \right)_{\nu}$.
Using the values of the neutrino luminosities,
mean energies and net rates for the wind region, 
we expect $Y_{e,{\rm eq}} \approx 0.42$ \citep[see, for example, eq. (77) of ][]{Qian1996} 
and $0.042 \, {\rm s} \la t_{\rm weak} \la 0.090 \, {\rm s}$. 
If we keep in mind that the absorption of neutrinos becomes less efficient
as the distance from the neutrino surfaces increases, we understand the presence of
both radial and vertical gradients for $Y_e$ inside the wind:
the early expanding matter has not enough time to reach $Y_{e,{\rm eq}}$, especially if it is initially
located at large distances from the relevant neutrino surfaces ($R_{\rm cyl} \ga 100 \, {\rm km}$).
On the other hand, matter expanding from the innermost part of the disc and moving in the funnel 
(within a polar angle $\la 40^o$), as well as matter that orbits several times around the HMNS before being
accelerated in the wind, increases its $Y_e$ close to the equilibrium value, but on a longer time-scale.\\
Also the matter entropy in the wind rises due to neutrino absorption. Typical initial values in the disc
are $s \sim 5 - 10 \, k_{\rm B} \, {\rm baryon^{-1}}$, while later we observe 
$s \sim 15-20 \, k_{\rm B} \, {\rm baryon^{-1}}$. The entropy is usually larger where the 
absorption is more intense and $Y_e$ has increased more. However, differently from $Y_e$, 
its spatial distribution is more uniform. Once the distance from the HMNS and the disc
has increased above $\sim 400 \, {\rm km}$, neutrino absorption becomes negligible and 
the entropy and the electron fraction are simply advected inside the wind.\\
The radial velocity in the wind increases from a few times $10^{-2} \, c$, just above the disc, to
a typical asymptotic expansion velocity of $0.08-0.09 \, c$. This acceleration is caused
by the continuous pressure gradient provided by newly expanding layers of matter.

\begin{figure*}
\begin{center}
\includegraphics[width = \linewidth]{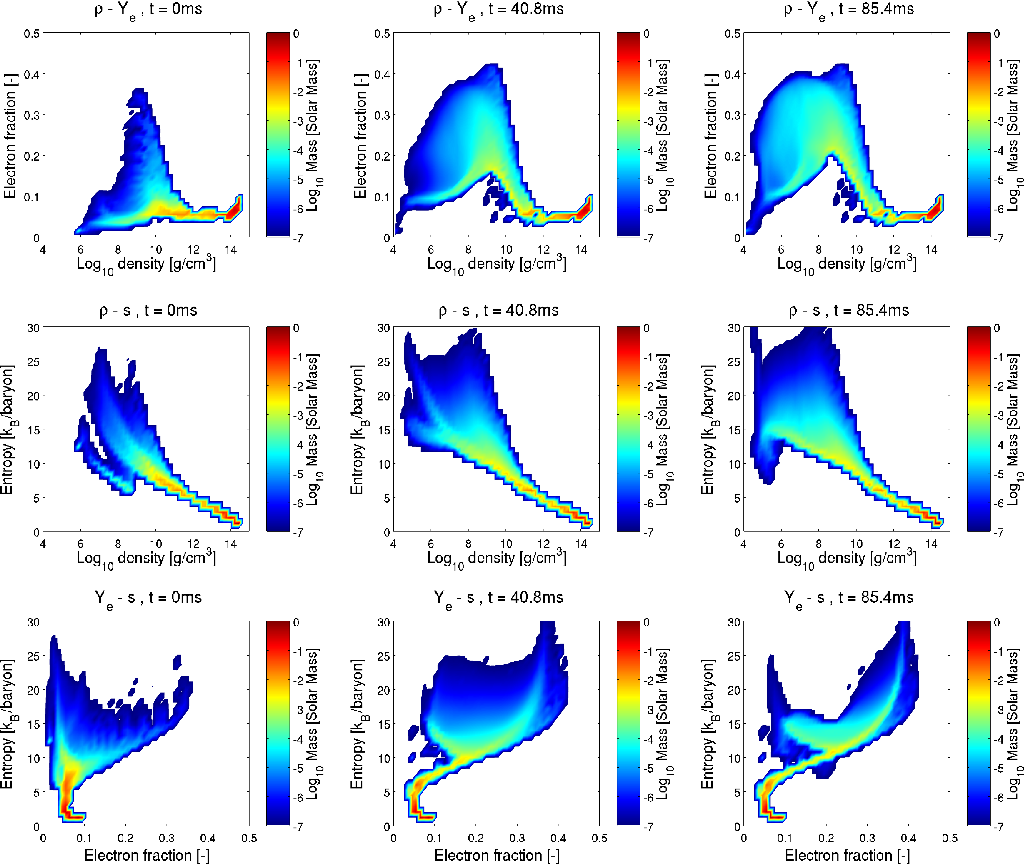}
\end{center}
\caption{Occurrence diagrams for $(\rho,Y_e)$ (top panel), $(\rho,s)$ (middle panels) 
and $(Y_e,s)$ (bottom panels), for the thermodynamical properties of matter in the whole system, 
at $t \approx 0 \, {\rm ms}$ (left column), $t \approx 40 \, {\rm ms}$ (central column) and
$t \approx 85 \, {\rm ms}$ (right column) after the beginning of the simulation.
Colour coded is a measure of the amount of matter experiencing specific thermodynamical conditions
inside the whole system. Occurrence smaller than $10^{-7}$ \msun have been omitted from the plot.}
\label{fig: occurrence}
\end{figure*}

To characterise the matter  properties, we plot in \reffig{fig: occurrence} \emph{occurrence diagrams} for couples of quantities, namely 
$\rho - Y_e$ (top row), $\rho - s$ (central row) 
and $ Y_e - s $ (bottom row), at three different times ($t = 0,40,85$ ms).
Colour coded is a measure of the amount of matter experiencing specific thermodynamical conditions
inside the whole system, at a certain time 
\footnote{A formal definition of the plotted quantity can be found in Sec. 2 of \cite{Bacca2012}.
However, in this work we don't calculate the time average.}.\\
We notice that most of the matter is extremely dense 
($\rho > 10^{11}\, {\rm g \, cm^{-3}}$), neutron rich ($Y_e < 0.1$) 
and, despite the large temperatures ($T > 1 \, {\rm MeV}$), at relatively 
low entropy ($s < 7 \, {k_B \, {\rm baryon}^{-1}}$). This matter correspond to the 
HMNS and to the innermost part of the disc, where matter conditions
change only on the long neutrino diffusion timescale, \refeq{eqn: diffusion time HMNS}, or
on the disc lifetime, \refeq{eqn: viscosity time}.
In the low density part of the diagrams ($\rho < 10^{11}\, {\rm g \, cm^{-3}}$), 
the expansion of the disc and the development of the wind can be traced.

\begin{figure*}
\begin{center}
\includegraphics[width = \linewidth]{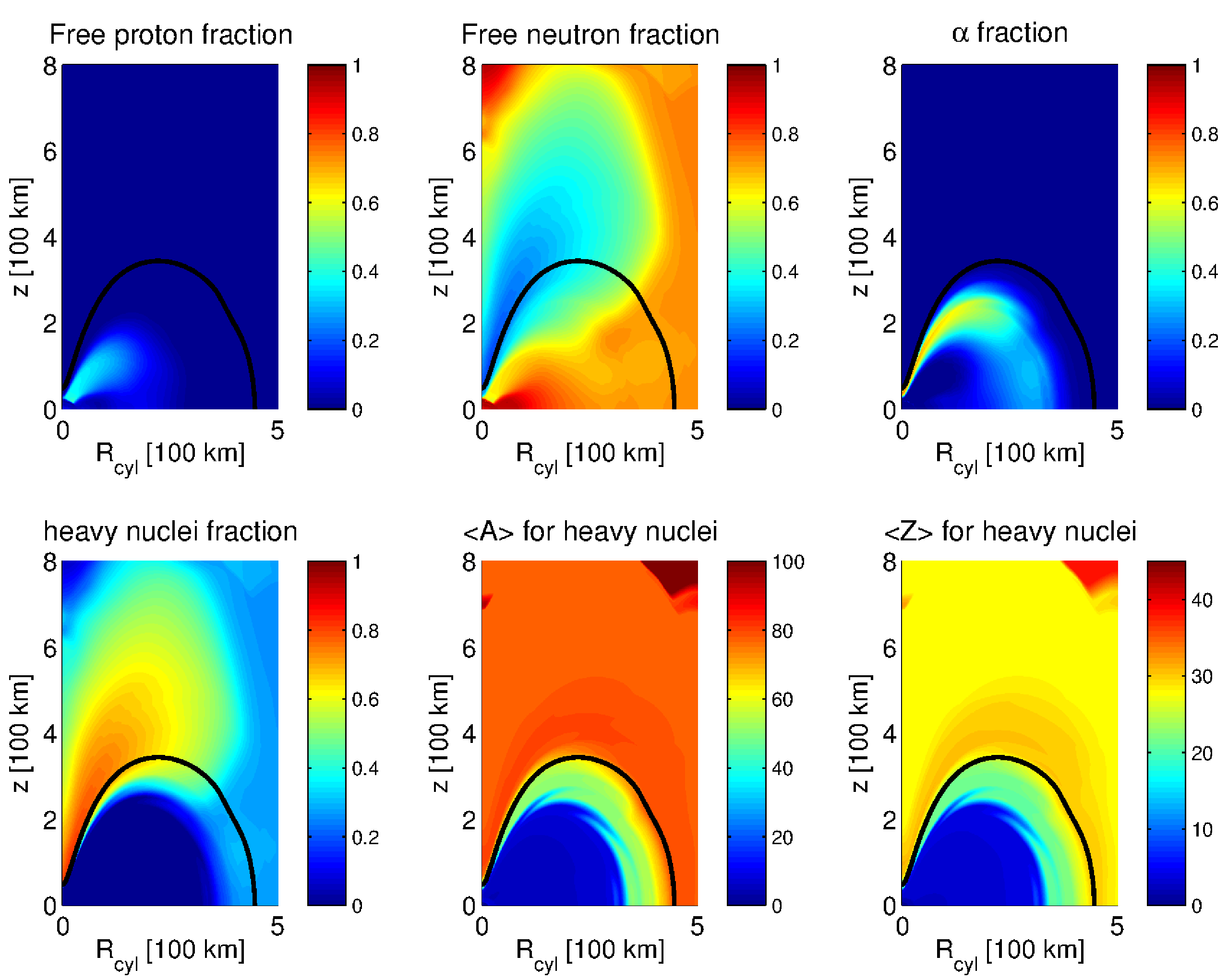}
\end{center}
\caption{Nuclear composition provided by the EoS (assuming everywhere NSE) in the disc and in the wind, at $t \approx 40 \, {\rm ms}$. 
On the top row, free proton (left), free neutron (centre) and
$\alpha$ particles (right) mass fractions. On the bottom row, heavy nuclei mass fraction (left), 
and mass number (centre) and atomic number (right) of the representative heavy nucleus. The black line represents the 
$T=0.5 \, {\rm MeV}$ surface.}
\label{fig: composition 40ms}
\end{figure*}

In \reffig{fig: composition 40ms} (a-d), we represent the mass fractions 
of the nuclear species provided by the nuclear EoS 
inside the disc and the wind, at 40 ms after the beginning of the simulation. 
Close to the equatorial plane ($|z| < 100 \, {\rm km}$), the composition 
is dominated by free neutrons. 
In the wind, the increase of the electron fraction corresponds to the conversion of 
neutrons into protons due to $\nue$ absorption. 
In the early expansion phase, the relatively high temperature 
($T \gg 0.6 \, {\rm MeV}$) 
favours the presence of free protons.
When the decrease of temperature allows the formation of nuclei, 
protons cluster into $\alpha$ particles and,
later, into neutron-rich nuclei. 
Then, the composition in the 
wind, at large distances from the disc, is distributed between free neutrons 
($0.4 \la X_n \la 0.6$) and heavy nuclei 
($0.6 \ga X_h \ga 0.4$, respectively).
The heavy nuclei component is described in the EoS 
by a representative average nucleus, assuming Nuclear Statistical Equilibrium (NSE) everywhere. 
In \reffig{fig: composition 40ms} (e-f), we have represented the values of its
mass and charge number.
The most representative nucleus in the wind corresponds often to ${}^{78}{\rm Ni}$.
The black line defines the surface across which 
the freeze-out from NSE is expected to occur 
($T = 0.5 \, {\rm MeV}$).
Outside it the actual composition will differ from the NSE prediction
(see \refsec{sec:discussion}).

%
%
\subsection{Ejecta}
\label{sec: ejecta}

Matter in the wind can gain enough energy from the neutrino absorption 
and from the subsequent disc dynamics to become unbound.
The amount of ejected matter is calculated as 
volume integral of the density and fulfils three criteria:
1) has positive radial velocity; 
2) has positive specific total energy; 
3) lies inside one of the two cones of opening angle $60^\circ$, 
vertex in the centre of the HMNS and axes coincident with the disc rotation axes.
The latter geometrical constraint excludes possible contributions coming from equatorial ejecta,
which have not been followed properly during their expansion.
The profile of $Y_e$ at the end of the simulation (see, for example, \reffig{fig: wind 85ms}) suggests
to further distinguish between two zones inside each cone, one at high (H: $0^\circ \leq \theta < 40^\circ$, where $\theta$ is
the polar angle) and one at low (L: $40^\circ \leq \theta < 60^\circ$) latitudes.\\
The specific total internal energy is calculated as:
\be
 e_{\rm tot} = e_{\rm int} + e_{\rm grav} + e_{\rm kin}.
 \label{eqn: total specific energy}
\ee
$e_{\rm grav}$ is the Newtonian gravitational potential, 
and $e_{\rm kin}$ is the specific kinetic energy. 
The specific internal energy $e_{\rm int}$ takes into account the nuclear recombination energy and,
to compute it, we use the composition provided by the EoS. For the nuclear binding energy
of the representative heavy nucleus, we use the semi-empirical nuclear mass formula
\citep[see, for example, the fitting to experimental nuclei masses reported by][]{Rohlf1994}:
in the wind, for $\langle A \rangle \approx 78$ 
and $\langle Z \rangle \approx 28$, the nuclear binding 
energy is $\sim 8.1 \, {\rm MeV \, baryon^{-1}}$.

\begin{figure*}
\begin{center}
\includegraphics[width = \linewidth]{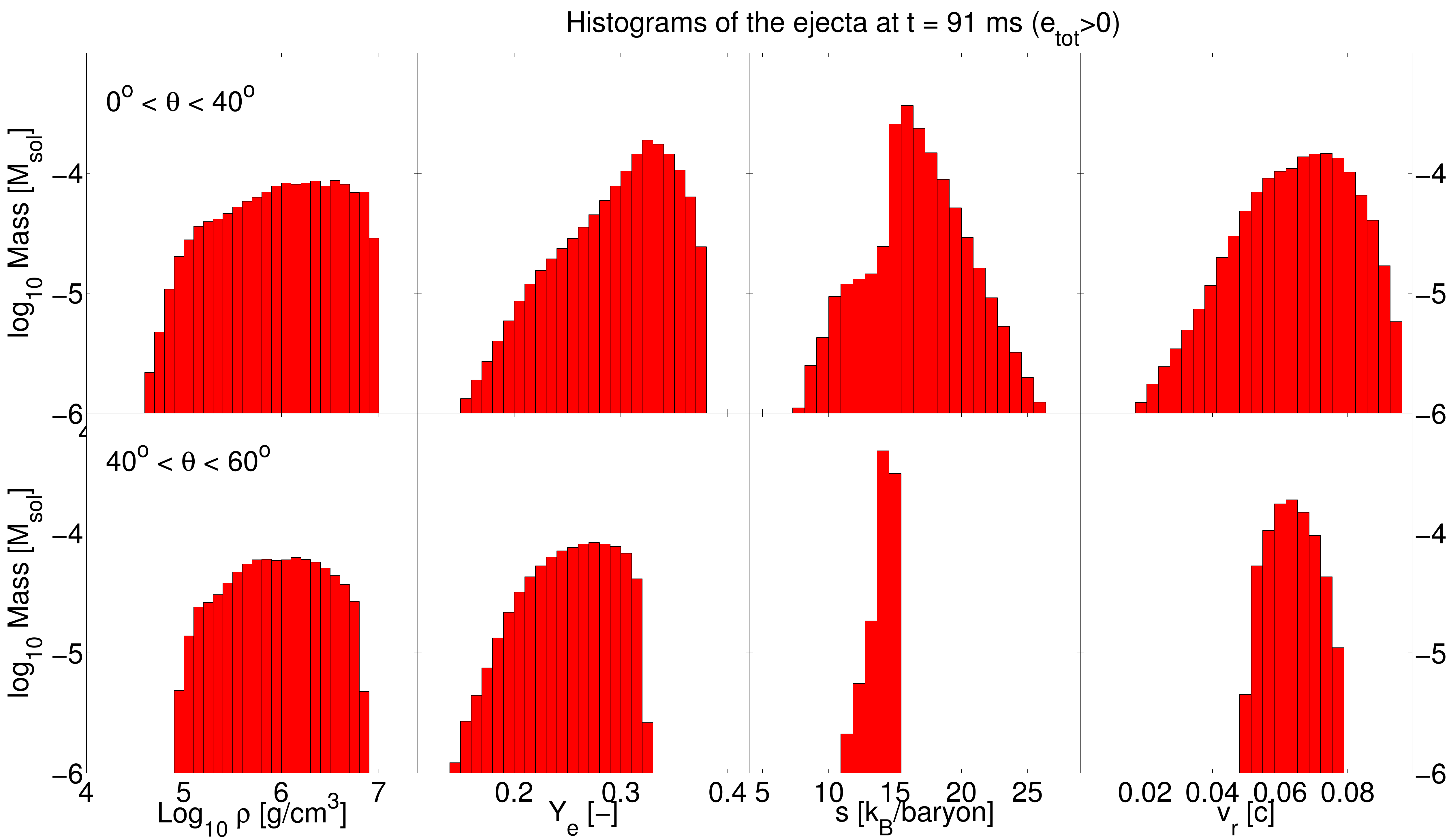}
\end{center}
\caption{Distributions in the $\nu$-driven wind ejecta binned by different physical properties. 
The different columns refer to 
density ($\rho$, left), 
electron fraction ($Y_e$, central-left), 
entropy per baryon ($s$, central-right) 
and radial velocity ($v_r$, right).
The top (bottom) panels refer to high (low) latitudes.}
\label{fig: histograms}
\end{figure*}

At the end of the simulation, $M_{\rm ej}(t = 91 \, {\rm ms}) \approx 2.12 \times 10^{-3}$ \msun,
corresponding to $\sim 1.2$ per cent of the initial disc mass ($M_{\rm disc} \approx 0.17$ \msun).
This mass is distributed between $M_{\rm ej,H}(t = 91 \, {\rm ms}) \approx 1.3 \times 10^{-3}$ \msun
at high latitudes and $M_{\rm ej,L}(t = 91 \, {\rm ms}) \approx 0.8 \times 10^{-3}$ \msun
at low latitudes.
In \reffig{fig: histograms}, we represent the mass distributions of density, electron fraction, entropy 
and radial velocity, for the ejecta at the end of our simulation.
At high latitude, the larger $\nue$ absorption enhances the electron fraction and the entropy more
than at lower latitudes. The corresponding mass distributions are broader, with peaks at
$Y_e \sim 0.31-0.35$ and $s \sim 15-20 k_{\rm B}/{\rm baryon}$. 
At lower latitudes, the electron fraction
presents a  relatively uniform distribution between 0.23 and 0.31, while the entropy has a very narrow peak
around 14-15 $k_{\rm B}/{\rm baryon}$. The larger energy and momentum depositions produce a faster
expansion of the wind close to the poles. This effect is visible in the larger average value and in
the broader distribution of the radial velocity that
characterises the high latitude ejecta.\\
To quantify the uncertainties in the determination of the 
ejecta mass, we repeat the previous calculation assuming an error of $ 0.5 \, {\rm MeV}$ in the estimate
of the nuclear recombination energy. For $M_{\rm ej,H}$ this translates in an uncertainty of
$\approx 7$ per cent, while in the case of $M_{\rm ej,L}$ the potential error is much larger ($\sim 50$ per cent).
This is a consequence of the different ejecta properties. At high latitudes, most of the free neutrons have
been incorporated into heavy nuclei, releasing the corresponding binding energy. Moreover, the large 
radial velocities ($v_r \sim 0.08-0.09 \, c$) provides most of the energy needed to overcome 
the gravitational potential.
At lower latitudes, the more abundant free neutrons and the lower radial velocities
($v_r \sim 0.06-0.07 \, c$) translate into a smaller ejecta amount, with a larger dependence on the 
nuclear recombination energy. Generally, we consider our numbers for the wind ejecta as lower limits, 
since a) we ignore the presence and likely amplification of magnetic fields which could substantially enhance the mass loss
\citep{Thompson2003a}, b) so far, we ignore heating from neutrino-annihilation and c) we do  not consider colatitudes $> 60^\circ$.

%
%

\section{Discussion}
\label{sec:discussion}

%
%
\subsection{Comparison with Previous Works}
The hierarchies we have obtained for the neutrino luminosities and mean energies agree 
with previous studies on the neutrino emission from neutron star mergers and their aftermaths.
In the case of Newtonian simulations, the compatibility is good
also from a quantitative point of
view, usually within 25 per cent
\citep[see, for example, the values obtained in ][ for the run H, to be compared with our
cooling luminosities]{Rosswog2013}.
On the other hand, general relativistic simulations 
(usually, limited in time to the first tens of milliseconds after the merger) 
obtain larger neutrino luminosities (up to a factor 2 or 3), due to larger
matter temperatures and stronger shocks
\citep[see, for example, ][]{Sekiguchi2011,Kiuchi2012,Neilsen2014}.
The higher temperatures reduce also the ratio between $\nueb$ and $\nue$ luminosities,
since the difference between charged current reactions on neutrons and protons diminishes ($k_{\rm B}T \gg Q$), 
and thermal pair processes are enhanced.

\cite{Dessart2009} studied the formation of the neutrino-driven wind, starting from 
initial conditions very similar to ours, in axisymmetric simulations that employ a multi-group
flux limited diffusion scheme for neutrinos. 
Our results agree with theirs concerning typical values of the neutrino luminosities and mean energies (with 
the exception of $\langle E_{\numt} \rangle$, see \refsec{sec: neutrino emission}), 
as well as their angular distributions. 
Also the shape and the extension of the neutrino surfaces inside the 
disc are comparable. 
There are, however, some differences in the temporal evolution:
while we observe almost stationary profiles, decreasing on a time-scale comparable with the expected disc
lifetime, their luminosities decrease faster. Also the difference between $\nue$ and $\nueb$ luminosities decreases,
leading to $L_{\nue} \approx L_{\nueb}$.
Both these differences can depend on the different accretion histories: the usage of the three dimensional
initial data (without performing axisymmetric averages) preserves all the initial local perturbations
and favours a substantial $\dot{M}$ inside the disc.\\
The removal and the deposition of energy, operated by neutrinos, is similar in the two cases. As a result,
the subsequent disc and wind dynamics agree well with each other.
The amount of ejecta and its electron fraction, on the other hand, show substantial differences: 
at $t \sim 100 \, {\rm ms}$, we observe a larger amount of unbound matter, whose electron fraction has
significantly increased.

The evolution of a purely Keplerian disc around a HMNS, under the influence of $\alpha$-viscosity and
neutrino self-irradiation, as a function of the lifetime of the central object, has been more 
recently investigated by \cite{Metzger2014}. 
They employ an axisymmetric HD model, coupled with a grey leakage scheme and a light bulb boundary luminosity
for the HMNS. They evolve their system for several seconds 
to study the development of the neutrino-driven wind and of the viscous ejecta.
In the case of a long-lived HMNS ($t_{ns} \ga 100 \, {\rm ms}$), 
our results for the wind are qualitatively similar to their findings:
we both distinguish between a polar outflow,
characterised by larger electron fractions, entropies and expansion time-scales, and a
more neutron rich equatorial outflow. The polar ejecta, mainly driven by neutrino absorption, represent
a meaningful, but small fraction of the initial mass of the disc (a few percent).
Quantitative differences, connected with the different initial 
conditions and the different neutrino treatment, are however present: their entropies and electron fractions 
are usually larger, especially at polar latitudes.

The importance of neutrinos in neutron star mergers 
has been recently addressed also by \cite{Wanajo2014}. They have shown that 
the inclusion of both neutrino emission and
absorption can increase the ejecta $Y_e$ to a wide range of values (0.1-0.4), leading to the production
of all the r-process nuclides from the dynamical ejecta. 
However, a direct comparison with our work is difficult since 1) their 
simulation employs a softer EoS, that amplifies general relativistic effects, and 2) their analysis is
limited to the dynamical ejecta and the influence of neutrinos on it during the first milliseconds after the 
merger.

%
%
\subsection{Nucleosynthesis in neutrino-driven winds}
\label{sub:nucleo}

\begin{table}
 \begin{center}
  \begin{tabular}{l|c|c|c|c|c|c|c}
  \hline
  Tracer & $Y_e$ 
         & s $[k_{\rm B} \ {\rm baryon}]$ 
         & $\langle A\rangle_{\rm final}$ 
         & $\langle Z\rangle_{\rm final}$
         & $X_{\rm La,Ac}$\\
  \hline
  L1  & 0.213 & 12.46 & 118.0 & 46.2 & $0.04$\\
  L2  & 0.232 & 11.84 & 107.1 & 42.5 & $0.009$\\
  L3  & 0.253 & 12.68 &  98.0 & 39.2 & $7\cdot10^{-5}$\\
  L4  & 0.275 & 12.73 &  90.2 & 36.4 & $1\cdot10^{-7}$\\
  L5  & 0.315 & 13.68 &  81.7 & 33.0 & $3\cdot10^{-12}$ \\
  \hline 
  H1  & 0.273 & 13.57 &  93.0 & 37.4 & $8\cdot10^{-7}$\\
  H2  & 0.308 & 14.69 &  83.3 & 33.7 & $6\cdot10^{-11}$\\
  H3  & 0.338 & 15.36 &  79.4 & 32.1 & $< 10^{-12}$\\
  H4  & 0.353 & 16.40 &  78.4 & 31.7 & $< 10^{-12}$\\
  H5  & 0.373 & 18.35 &  76.8 & 31.0 & $< 10^{-12}$\\
  \hline
  \end{tabular}
 \end{center}
 \caption{Parameters of representative tracers and corresponding
 nucleosynthesis: electron fraction $Y_e$, specific entropy per baryon $s$,
 average atomic mass $\langle A\rangle_{\rm final}$ and electric charge 
 $\langle Z\rangle_{\rm final}$ of the resulting nuclei, and the total mass
 fractions of Lanthanides and Actinides in the resulting nucleosynthetic mix.
 The latter are important for estimating opacities at the location of the
 tracers.
 }
 \label{tab:tracers}
\end{table}

\begin{figure}
  \includegraphics[width=0.50 \textwidth]{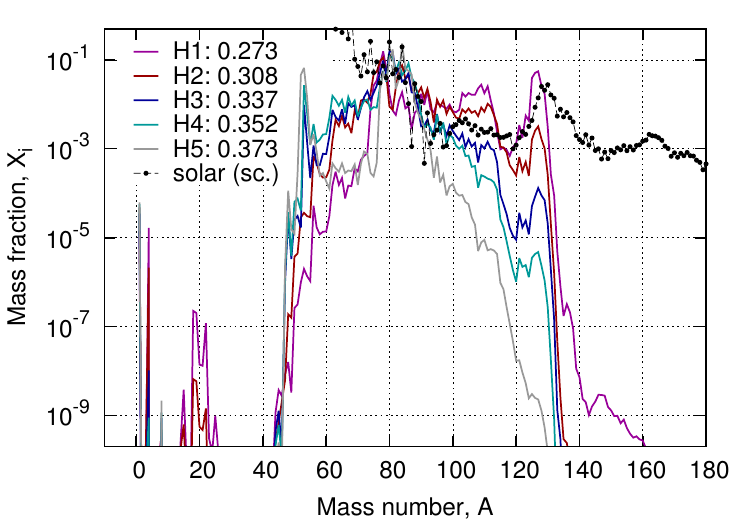} \\
  \includegraphics[width=0.50 \textwidth]{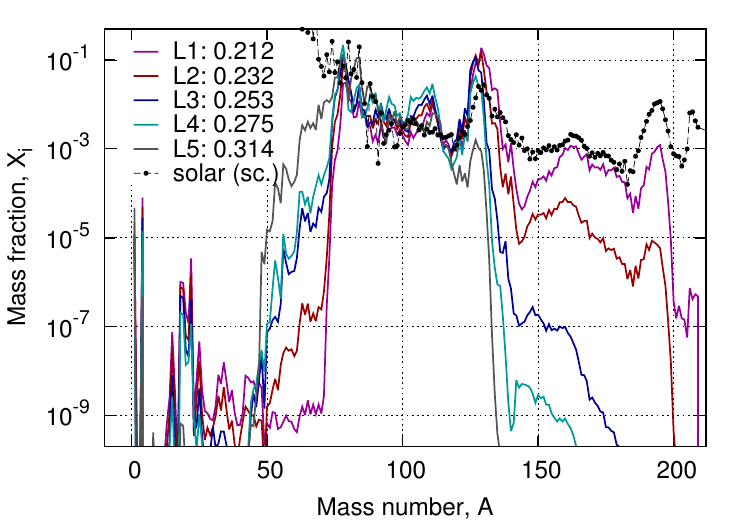}
\caption{Summed final mass fractions for representative tracers.
Top and bottom panels correspond to high-latitude (H1-H5) and low-latitude
(L1-L5) tracers, respectively (see~\reftab{tab:tracers} for parameters of
individual tracers). Solar r-process abundances (scaled) are also shown for
comparison.}
\label{fig:nucleosynthesis}
\end{figure}

During our simulation, we have computed
trajectories of representative tracer particles (Lagrangian particles, passively 
advected in the fluid during the simulation).
The related full nucleosynthesis will be explored in more detail in future work.
To get a first idea about the possible nucleosynthetic signatures, we have selected ten tracers, 
extrapolated and post-processed with a nuclear network.
These tracers are equally distributed between the high and the low latitude
region (5+5). Inside each region, we have picked the particles that represent
the most abundant conditions in terms of entropy and electron fraction in the
ejecta at $t \approx 90 \, {\rm ms}$.
\reftab{tab:tracers} lists parameters of the selected tracers.

For the nucleosynthesis calculations we employ the WinNet nuclear reaction
network~\citep{Winteler2012a,Winteler2012}, which represents an update of BasNet
network code~\citep{Thielemann2011}. 
The ingredients for the network that we use are the same as described
in~\cite{Korobkin2012}. We have also included the feedback of nuclear heating
on the temperature, but we ignore its impact on the density, since previous
studies have demonstrated that for the purposes of nucleosynthesis this impact
can be neglected~\citep{Rosswog2014a}.
In this exploratory study, we also do not include neutrino irradiation.
Instead we use the final value of electron fraction from the tracer to initialise
the network. In this way, we effectively take into account the final neutrino
absorptions. Our preliminary experiments show that neutrino irradiation has an
effect equivalent to vary $Y_e$ by a few percent, which is a
correction that will be addressed in future work.
It is also worth mentioning that the situation is even less simple if one
takes into account neutrino flavour oscillations, which may alter the
composition of the irradiating fluxes significantly, depending on the
densities and distances involved~\citep{Malkus2014}.

\reffig{fig:nucleosynthesis} shows the resulting nucleosynthetic mass fractions, 
summed up for different atomic masses, and \reftab{tab:tracers} lists the
averaged properties of the resulting nuclei.
As expected, lower electron fractions lead to an r-process
with heavier elements, and for the lowest values of $Y_e$ even the elements up to
the third r-process peak ($A\sim190$) can be synthesised.
However, due the high sensitivity to the electron fraction, wind
nucleosynthesis cannot be responsible for the observed astrophysical robust
pattern of abundances of the main r-process elements. 
On the other hand, it could successfully contribute to the weak r-process in
the range of atomic masses from the first to second peak 
($70 \la A \la 110$).

\reffig{fig:nucleosynthesis} also illustrates that heavier elements tend to
be synthesised at lower latitudes, closer to the equatorial plane. This has
important consequences for directional observability of associated electromagnetic
transients. Material, contaminated with Lanthanides or Actinides is expected
to have opacities that are orders of magnitudes larger than those of iron group 
elements. Therefore, the corresponding electromagnetic signal is expected to
peak in the infrared.
\cite{Kasen2013} estimates that as little as $X_{\rm La,Ac} \ga 0.01$ per cent of
these ``opacity polluters'' could be enough to raise the opacities by a factor
of hundred.
\reftab{tab:tracers} lists also the computed mass fraction of the opacity
polluters, which turns out to be negligible for high-latitude tracers, while
being quite significant for low-latitude ones.
We therefore expect that the signal from the wind outflow will look much
redder, dimmer and peak later if the outflow is seen from equatorial rather
than polar direction.
Additionally, if seen from the low latitudes, the signal from the wind outflow
can be further obscured by the dynamical ejecta. 
Thus, for the on-axis orientation the signal has better prospects of
detection, therefore making follow-up observations of short GRBs more
promising.
We will discuss these questions in detail in~\refsec{sub:macronova} below. 

%
%
\subsection{Electromagnetic transients}
\label{sub:macronova}

\begin{figure*}
\begin{tabular}{cc}
  \includegraphics[width=0.46\textwidth]{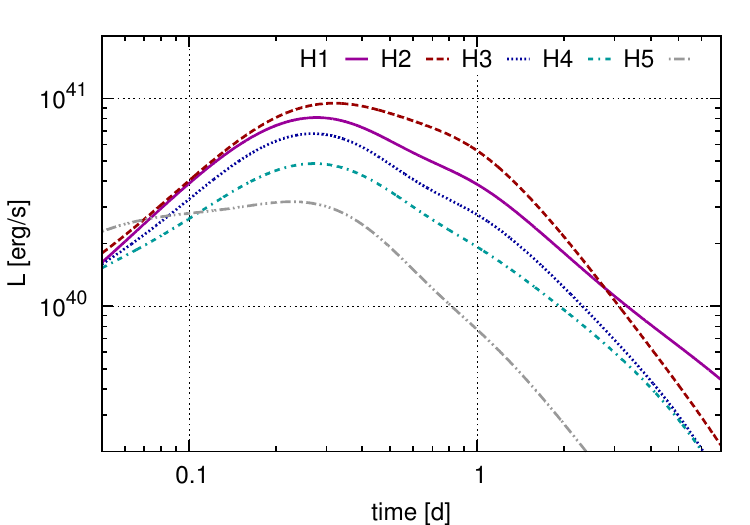} &
  \includegraphics[width=0.46\textwidth]{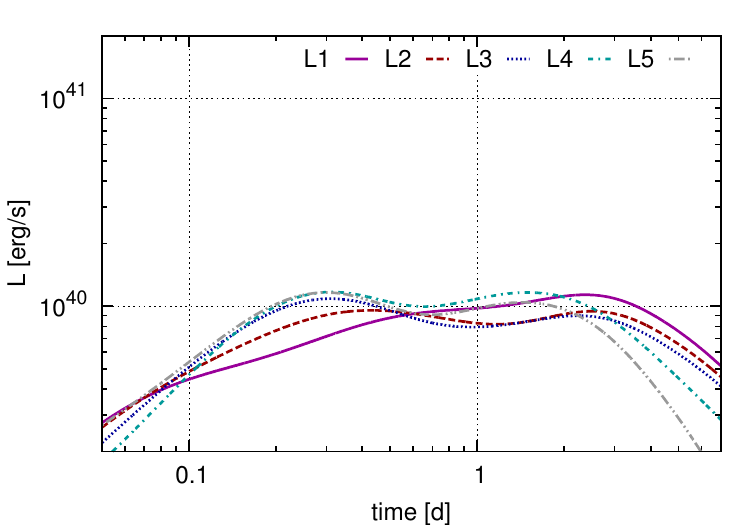}
  \\
  \includegraphics[width=0.46\textwidth]{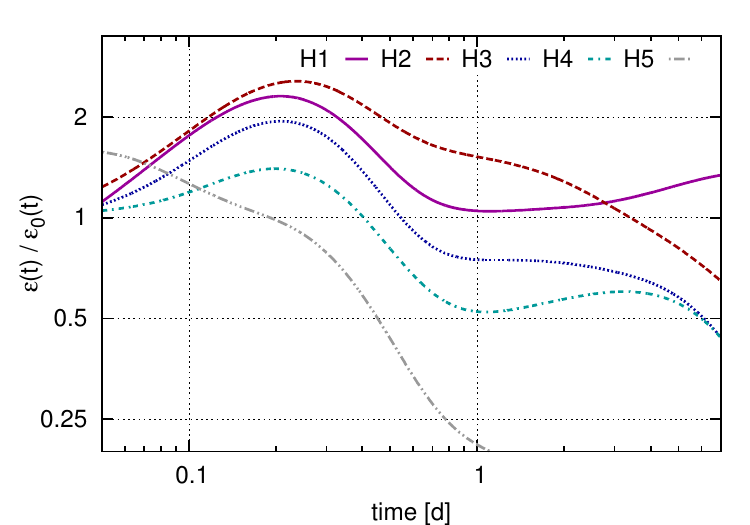} &
  \includegraphics[width=0.46\textwidth]{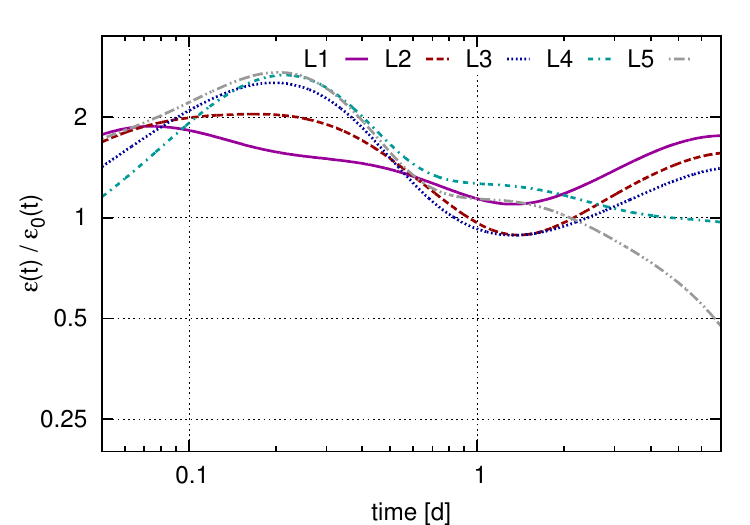}
  \\
  \includegraphics[width=0.46\textwidth]{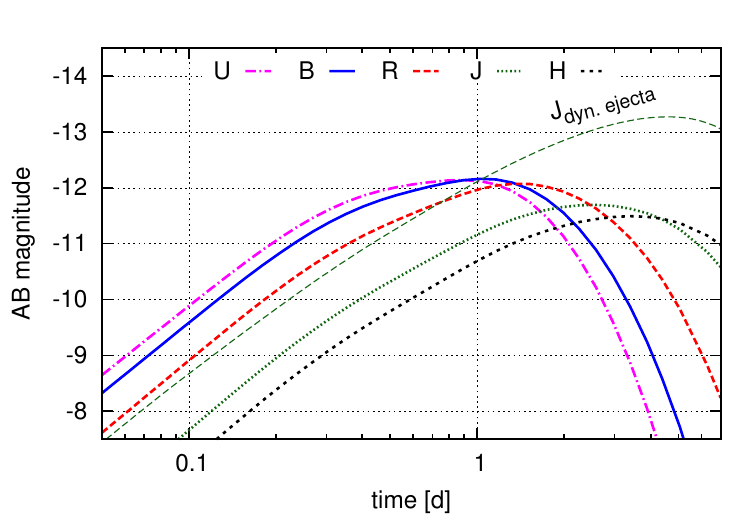} &
  \includegraphics[width=0.46\textwidth]{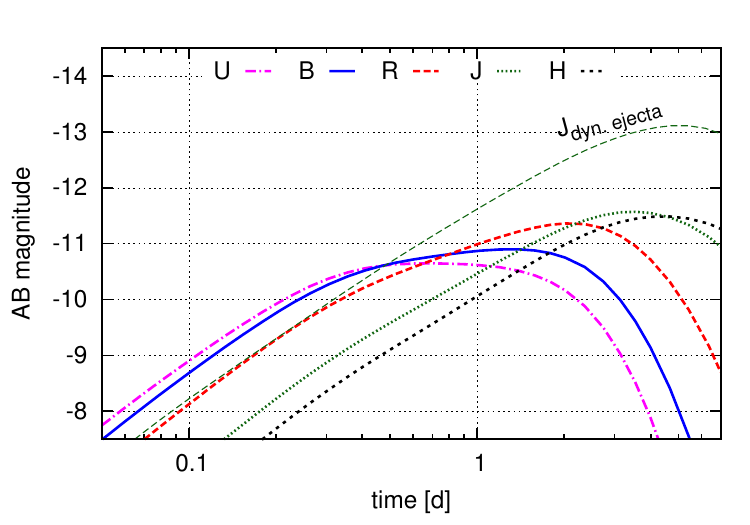}
\end{tabular}  
\caption{Electromagnetic transients due to the radioactive material produced in the neutrino-driven wind. 
The left column refers to material ejected at high latitudes (H1-H5),
the right column shows the results for the low latitudes (L1-L5).
Top row: predicted macronova lightcurves (bolometric luminosity), 
calculated with uniform-composition spherically symmetric Kulkarni-type
models. Model parameters: ejected mass $
m_{\rm ej}=2\cdot10^{-3}~{\rm M_\odot}$, 
expansion velocity $v_{\rm e}=0.08\;c$.
Opacity is taken to be $1~{\rm cm}^2/{\rm g}$
and $10~{\rm cm}^2/{\rm g}$ for high- and low-latitude tracers respectively.
Middle row: radioactive heating rate for the representative tracers, normalised 
to $\dot{\epsilon}_0=10^{10}t_d^{-1.3}~{\rm erg}/({\rm g}\cdot{\rm s})$. 
Bottom row: broadband AB magnitudes in five different bands, calculated for
the case when the wind outflow is viewed from the 'top' (left panel) and
'side' (right panel). For comparison, the J band signal from the dynamic ejecta is
superimposed.}
\label{fig:macronovae}
\end{figure*}

In \refsec{sec: ejecta}, we have estimated the amount of mass ejected 
at the end of our simulation ($M_{\rm ej} (t \approx 90 \,{\rm ms}) \approx 2.12 \times 10^{-3}$ \msun ).
As discussed there, it needs to be considered as a lower
limit on the mass loss at that time. The neutrino emission, however, will continue beyond that time and keep driving the
wind outflow. We make here an effort to estimate the total mass loss caused by neutrino-driven winds during the disc lifetime.
During our simulation, the temporal evolution of the accretion rate on the HMNS 
(\reffig{fig: accretion_eta_time}) is well described by
\begin{equation}
\dot{M}(t) \approx 0.76 \, \exp{\left[ - t/(0.124 \, {\rm s}) \right]} \, \Ms \, {\rm s^{-1}}.
\end{equation}
We notice that, according to this expression, the total accreted mass is
smaller than the initial mass of the disc:
\begin{equation}
M_{\rm acc} \equiv \int_0^{\infty} \dot{M} \, {\rm d}t \approx 0.095 \Ms < M_{\rm disc}(t=0) \approx 0.17 \Ms.
\end{equation}
This discrepancy can be interpreted as the effect of the wind outflow and of
the disc evaporation. The beginning of the latter process has already been
observed in our model, but not followed properly due to computational
limitations.
At $t \approx 0.285 \, {\rm s}$ the HMNS has accreted 90 per cent of $M_{\rm acc}$. This
agrees well with the viscous lifetime of the disc (\refeq{eqn: viscosity
time}), so we consider $t \approx 0.3 \, {\rm s}$ as a good estimate for the disc lifetime.
Since the wind is powered by neutrino absorption, we assume that the mass of the ejecta
is proportional to the energy emitted in neutrinos during the disc life
time:
\begin{equation}
M_{\rm ej}(t = 0.300 \, {\rm s}) = 
\frac{ \int_0^{0.300 \, {\rm s}} L_{\nu,{\rm cool}} \, {\rm d}t }{ \int_0^{0.090 \, {\rm s}} L_{\nu,{\rm cool}} \, {\rm d}t} 
\, M_{\rm ej}(t = 0.090 \, {\rm s}).
\end{equation}
To model $L_{\nu}(t)$ for $t>90 \, {\rm ms}$, we consider two possible cases: 
\begin{itemize}
    \item [A)]  the HMNS collapses after the disc has been completely accreted;
    \item [B)]  it collapses promptly at the end of our simulations. 
\end{itemize}
For both cases, we extrapolate linearly the luminosities from \reffig{fig: lum with time}.
But in case B, we decrease the neutrino luminosity by 50 per cent, to account for the
lack of contribution from the HMNS and the innermost part of the disc after
the collapse (see \refsec{sec: neutrino emission}).
Our final mass extrapolations are listed in \reftab{table: ejecta masses}. 
So in summary,  we find $4.87 \times 10^{-3}$ \msun for case A and
$3.49 \times 10^{-3}$ \msun for case B. Given that we consider
these numbers as lower limits, this implies that the wind would provide a substantial 
contribution to the total mass lost in a neutron star merger (and likely similar for a neutron star-black hole
merger; for an overview over the dynamic ejecta masses see \cite{Rosswog2013}).

\begin{table}
 \begin{center}
    \begin{tabular}{| c | c | c | c | c | c |}
    \hline
   Case  & $t [{\rm s}]$   & $t_{\rm ns}[{\rm s}]$ & $M_{\rm{ej,H}} [\Ms]$ & $M_{\rm{ej,L}} [\Ms]$ & $M_{\rm{ej}} [\Ms]$ \\ \hline \hline
   A/B   & $0.09$          & $\ge 0.09 \,{\rm s}$  & $1.29 \cdot 10^{-3}$  & $0.82\cdot 10^{-3}$   & $2.11\cdot 10^{-3}$ \\ 
         &                 &                       &                       &                               \\ \hline
   Case  & $t [{\rm s}]$   & $t_{\rm ns}[{\rm s}]$& $M_{\rm{ej,H}} [\Ms]$ & $M_{\rm{ej,L}} [\Ms]$ & $M_{\rm{ej}} [\Ms]$ \\ \hline \hline
   A     & $0.3$      & $> 0.3 \,{\rm s}$     & $2.98 \cdot 10^{-3}$  & $1.89\cdot 10^{-3}$   & $4.87\cdot 10^{-3}$ \\
   B     & $0.3$      & $\sim 0.09 \,{\rm s}$ & $2.13 \cdot 10^{-3}$  & $1.36\cdot 10^{-3}$   & $3.49\cdot 10^{-3}$ \\ 
   \end{tabular}
  \end{center}
\caption{Values of the calculated (top, for $t \approx 90 \, {\rm ms}$) and extrapolated 
        (bottom for $t \approx 300 \, {\rm ms}$) ejected masses, for the high (H) and low (L) latitude regions, and their sum.
        $t_{\rm ns}$ refers to the time-scale for the HMNS to collapse to a black hole.
   }
  \label{table: ejecta masses}
\end{table}

With these mass estimates, we compute expected lightcurves for each tracer,
using the semi analytic spherically-symmetric models of macronovae by
\cite{Kulkarni2005}, the same as the ones described
in \cite{Grossman2014}.
\reffig{fig:macronovae} shows the resulting lightcurves (top row) for the wind
outflow mass from the case A. 
Each lightcurve corresponds to a simplified case when the entire wind ejecta
evolves according to the thermodynamic conditions of one specific tracer.
In this work, we do not take into account spatial or temporal variation of the
electron fraction within the wind outflow, but we assume different opacities
for the high-latitude and low-latitude tracers. Motivated by recent work of 
\cite{Kasen2013} and confirmed by \cite{Tanaka2013a}, we
take a uniform grey opacity of $\kappa=10 \, {\rm cm}^2 \, {\rm g}^{-1}$
for the low-latitude tracers that have a low $Y_e$ and produce non-negligible
amounts of Lanthanides and Actinides. For the high-latitude, higher $Y_e$ tracers we use
$1\,{\rm cm}^2 \, {\rm g}^{-1}$.
The tracers result in a wide variety of potential lightcurves, whose shape
reflects individual nuclear heating conditions for a specific tracer. 
The middle row of \reffig{fig:macronovae} shows the individual heating rates,
normalised to the power law 
$\dot{\epsilon}_0=10^{10}t_d^{-1.3}~{\rm erg} \, {\rm g}^{-1} \, {\rm s}^{-1})$.
Differences in the shapes of the heating rates for different tracers
are due to the dominance of different radioactive elements 
at late times \citep{Grossman2014}.
Despite the variety of macronovae for different tracers, the actual lightcurve
will lie somewhere in between, and the individual differences in the heating
rates will be smoothed out.
The bottom row represents averaged broadband lightcurves from the
high-latitude (left panel) and low-latitude (right panel) wind ejecta. 
The high-latitude case shows a pronounced peak in the B band at 
$t\sim1.3\,{\rm d}$, while the higher opacities 
for the low-latitude tracers make the lightcurve
dimmer, redder and cause them to peak later.

\begin{figure*}
\begin{center}
  \begin{tabular}{cc}
    \includegraphics[width=0.46\textwidth]{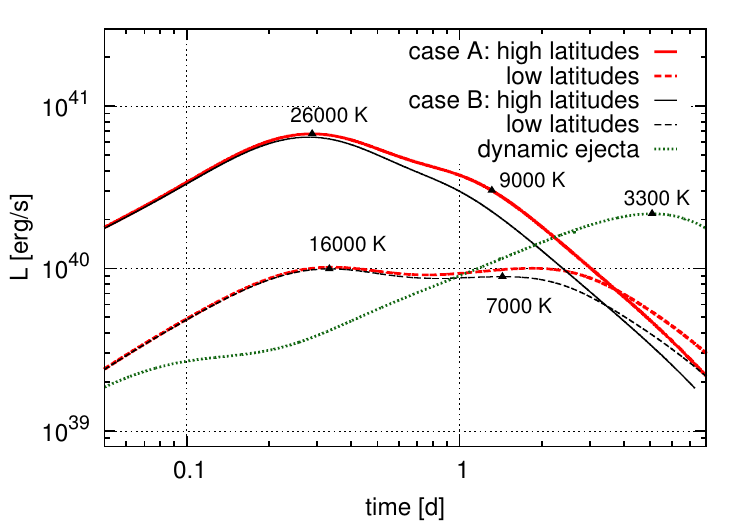}
    &
    \includegraphics[width=0.46\textwidth]{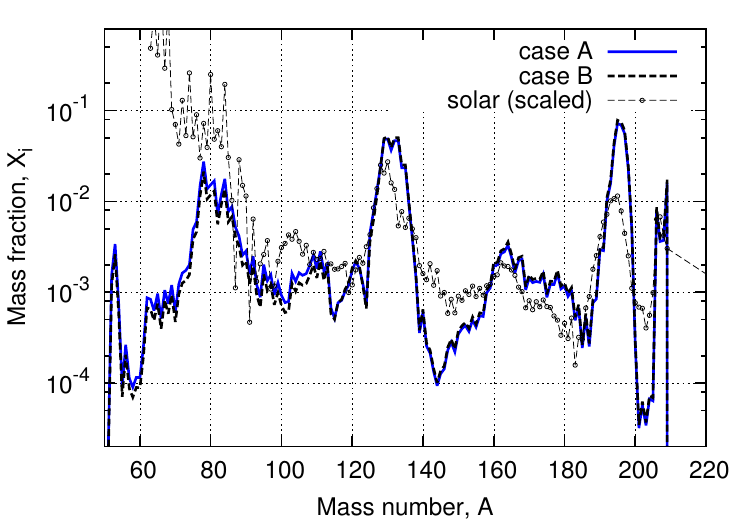}
  \end{tabular}  
\end{center}
\caption{Bolometric lightcurves (left) and summed abundances (right) for the
two cases of a long-lived (case A) and a short-lived (case B) HMNS.
The left panel shows separately bolometric lightcurves of low-latitude and
high-latitude outflows, as well as the lightcurve for the dynamical ejecta
from the same merger simulation. The plot also shows the effective
temperatures of the macronovae at the key points on the curves.
On the right panel, the wind abundances have been added to the abundances from
dynamical ejecta, for which we took the total ejected mass of 
$1.3\cdot10^{-3}$ \msun from the merger simulation.
}
\label{fig:casesAB}
\end{figure*}

An interesting question is whether or not the collapse time of the HMNS could 
possibly be inferred from the EM signal, assuming that the collapse happens after
the wind has formed ($t_{\rm ns} \ga 100 \, {\rm ms}$). Therefore we compare in
\reffig{fig:casesAB} (left panel)  the averaged bolometric
lightcurves for the cases A and B of long- and short-lived HMNS. 
The plot shows the low-latitude and high-latitude components
separately, as well as the lightcurve for the dynamic ejecta for the same merger case.
The two cases differ very little, mainly because the mass of the wind
component changes only by a factor of $\sim1.5$, and the lightcurve is not
very sensitive to this mass.
The long-lived HMNS (case A) is slightly brighter, but it is not likely that
the two cases can be discriminated observationally.
The difference between high- and low-latitude regions shows that perhaps
geometry of the outflow and its orientation relative to the observer plays
much more important role in the brightness and colour of the expected
electromagnetic signal.
Similarly, there is practically no difference in the total summed
nucleosynthetic yields for cases A and B (\reffig{fig:casesAB}, right panel).
Thus it may be difficult to extract the HMNS collapse time-scale from
the macronova signal.

%
%
\section{Conclusions}
\label{sec:conclusions}

We have explored the properties of the neutrino-driven wind
that forms in the aftermath of a BNS merger.
In particular, we have discussed their implications in terms of the r-process nucleosynthesis
and of the electromagnetic counterparts 
powered by the decay of radioactive elements in the expanding ejecta.

To model the wind, we have performed for the first time 3D Newtonian hydrodynamics 
simulations, covering an interval of $\approx 100 \, {\rm ms}$ after the merger,
and a radial distance of $\ga 1500 \, {\rm km}$ from the HMNS, with high spatial resolution 
inside the wind. Neutrino radiation has been treated by a computationally efficient, multi-flavour 
Advanced Spectral Leakage scheme, which includes consistent neutrino absorption rates in
optically thin conditions.
Our initial configuration is obtained from the direct re-mapping of the matter distribution 
of a 3D SPH simulation of the merger of two non-spinning 1.4 \msun neutron stars 
\citep[][and references therein]{Rosswog2007c}, at $\approx 15$~ms after the first contact.
The consistent dimensionality and the high compatibility between the two models 
do not require any global average nor any ad hoc assumption for the matter profiles inside the disc.\\

Our major findings are:
\begin{enumerate}
\item the wind provides a substantial contribution to the total
mass lost in a BNS merger.
At the end of our simulation ($\approx 100$ ms after the merger), we compute $2.12 \times 10^{-3}$ \msun~ 
of neutron-rich ($0.2 \la Y_e \la 0.4$) ejected matter, corresponding to 1.2 per cent of the 
initial mass of the disc. We distinguish between a high-latitude ($50^{\circ} - 90^{\circ}$) 
and a low-latitude ($30^{\circ} - 50^{\circ}$) component
of the ejecta. The former is subject to a more intense neutrino irradiation and 
is characterised by larger $Y_e$, entropies and expansion velocities.
We estimate that, on the longer disc lifetime, the ejected mass can increase to
$3.49-4.87 \times 10^{-3}$ \msun, where the smaller (larger) value refers to
a quick (late) HMNS collapse after the end of our simulation.
\item The tendency of $Y_e$ to increase with time above 0.3, especially at high latitudes,
suggests a relevant contribution to the nucleosynthesis of the weak r-process elements from the wind, 
in the range of atomic masses from the 
first to the second peak. Matter ejected closer to the disc plane retains a lower
electron fraction (between 0.2 and 0.3), and produces nuclei from the first to the third peak, 
without providing a robust r-process pattern.
\item The geometry of the outflow and its orientation relative to the observer have an important
role for the properties of the electromagnetic transient.
According to our results, the high-latitude outflow can power a
bluer and brighter lightcurve, that peaks within one day after the merger. 
Due to the partial contamination of Lanthanides and 
Actinides, the low-latitude ejecta is expected to have higher opacity and to peak later, 
with a dimmer and redder lightcurve.
\item A significant fraction of the neutrino luminosity is provided by the accretion
process inside the disc. This fraction is expected to power a (less intense) baryonic wind also if
the HMNS collapses to a BH before the disc consumption. According to our calculations, the
collapse time-scale has a minor impact on the possible observables (electromagnetic counterparts and
nucleosynthesis yields), at least if the collapse happens after the wind has formed 
and weak equilibrium had time to establish inside it. \cite{Metzger2014} indicate 
that more meaningful differences can be potentially seen, in the case of an earlier collapse.
This scenario requires further investigations.
\end{enumerate}

Our 3D results show a good qualitative agreement with the 2D results obtained by \cite{Dessart2009}
for a similar initial configuration, especially for the neutrino emission and the wind dynamics.
Meaningful quantitative differences are still present, probably related to the different accretion 
and luminosity histories  inside the disc. The distinction between a high-latitude and a low-latitude region in the ejecta
is qualitatively consistent with recent 2D findings of \cite{Metzger2014}.\\
The results we have found for the amount of wind ejecta has to be considered as lower limits, 
since in our model we ignore the effects of magnetic fields and neutrino-annihilation in optically thin conditions.
In particular, the latter is expected to deposit energy very efficiently in the funnel above the HMNS poles. 
The calculation of this energy deposition rate for our model and its implication 
for the sGRB mechanism will be discussed in a future work.\\
The wind ejecta has to be complemented with the dynamical ejecta
and with the outflow coming from the viscous evolution of the disc.
These other channels are expected to provide low-latitude outflows, with an electron fraction similar
or lower than the one obtained by the low-latitude wind component 
\citep[see, for example, ][ and references therein]{Rosswog2013,Metzger2014}. 
Instead, the high-latitude wind component seems to be peculiar in terms of outflow geometry, 
nucleosynthesis yields and related radioactively powered electromagnetic emission.\\

This work represents one of the first steps towards a physically consistent and complete model of the aftermath
of BNS mergers, including the effect of neutrino irradiation.
Our preliminary calculations regarding the nucleosynthesis and the electromagnetic
counterparts motivate further analysis and investigations.
Moreover, additional work has to be done to develop more accurate and complete radiation hydrodynamics treatments, 
to include other relevant physical ingredients (like magnetic fields and General Relativity),
and to explore the present uncertainties in terms of nuclear matter properties and neutrino physics.

%
%
\section*{Acknowledgements}
The authors thank F.K. Thielemann for useful discussions and for reading the manuscript.
AP and AA were supported by the Helmholtz-University Investigator grant No. VH-NG-825. 
The work of SR has been supported by the Swedish Research Council (VR) under grant 621-2012-4870.
RC and ML acknowledge the support from the HP2C Supernova project and the ERC grant FISH.
AP, SR and ML thank the MICRA-2009 Workshop and the Niels Bohr Institute for their hospitality during the summer of
2009, when this project started.
AP thanks the Jacobs University Bremen for its hospitality in February 2010, October 2010 and April 2012,
and the Stockholm University for its hospitality in June 2013. 
SR thanks the University of Basel for its hospitality in June 2010.
AP and SR thank COMPSTAR for the Short Visit Grants 3369 and 3536.
AP,RC RK and ML thank the use of computational resources provided by 
the Swiss SuperComputing Center (CSCS), under the allocation grants s414.
The SPH simulations for the project have been performed on the facilities
of the H\"{o}chstleistungsrechenzentrum Nord (HLRN).

%
%
\bibliographystyle{mn2e}

\bsp

\label{lastpage}

\end{document}